\documentclass[mlmain]{jmlr}

\usepackage{longtable}
\usepackage{booktabs}
\usepackage[load-configurations=version-1]{siunitx}
\usepackage{bm}
\usepackage{mathtools}
\usepackage{physics}
\usepackage{comment}

\theorembodyfont{\upshape}
\theoremheaderfont{\scshape}
\theorempostheader{:}
\theoremsep{\newline}

\usepackage{xcolor}
\usepackage[normalem]{ulem}

\makeatletter
\g@addto@macro\normalsize{%
  \setlength\abovedisplayskip{5pt plus 2pt minus 2pt}%
  \setlength\belowdisplayskip{5pt plus 2pt minus 2pt}%
  \setlength\abovedisplayshortskip{0pt plus 2pt}%
  \setlength\belowdisplayshortskip{3pt plus 2pt minus 1pt}}
\makeatother
\jmlrproceedings{PMLR}{Proceedings of Machine Learning Research}

\jmlrvolume{}
\firstpageno{1}
\jmlryear{2026}
\jmlrworkshop{Symmetry and Geometry in Neural Representations}

\title[Local Predictive Directions]{Self-Supervised Learning of Local Predictive Directions as a Neural Computation}

\newcommand{\authorbox}[3]{\parbox[t]{0.485\textwidth}{\raggedright\Name{#1}\\ {\small\mdseries\scshape #2}\\ \addr #3}}
\author{%
\authorbox{Marco Zenari\textsuperscript{*}}{marco.zenari.2@phd.unipd.it}{University of Padova, Padova, Italy}\hfill
\authorbox{Xuehao Ding\textsuperscript{*}}{xding@flatironinstitute.org}{Flatiron Institute, New York, NY, USA}%
\AND
\authorbox{Daniel Burnham}{dburnham@uoregon.edu}{University of Oregon, Eugene, OR, USA}\hfill
\authorbox{Austin R. {Ellis-Mohr}}{austine4@illinois.edu}{University of Illinois at Urbana-Champaign,\\ Urbana, IL, USA}%
\AND
\authorbox{Anirvan M. Sengupta}{anirvans@physics.rutgers.edu}{Rutgers University, New Brunswick, NJ, USA\\ Flatiron Institute, New York, NY, USA}\hfill
\authorbox{Dmitri B. Chklovskii}{dchklovskii@flatironinstitute.org}{Flatiron Institute, New York, NY, USA\\ Neuroscience Institute, NYU Langone Medical Center, New York, NY, USA}%
}

\begin{document}
\maketitle
\begingroup\renewcommand\thefootnote{*}\footnotetext{Equal contribution.}\endgroup
\begin{abstract}
Natural sensory dynamics are globally nonlinear but often exhibit simple local geometry, with nearby trajectories evolving preferentially along a small number of predictive directions. We ask how neural circuits could discover this geometry directly from temporal experience. We introduce Local Predictive Directions (LPD), a self-supervised framework that learns local directions whose projections maximally predict near-future trajectory evolution. LPD applies past–future canonical correlation analysis to temporal differences, making the learned representation invariant to translations of the underlying state while avoiding estimation of equilibria or a global dynamical model. To represent systems with multiple predictive regimes, we introduce a non-negative similarity-matching objective, with offline and online algorithms, that partitions trajectory segments into predictive domains, each associated with a distinct local predictive direction. The online algorithm admits a natural neural-network interpretation: the resulting activities provide soft domain assignments, while synaptic weights encode the corresponding local predictive filters. On stochastic nonlinear dynamical systems, LPD recovers the local predictive geometry and partitions trajectories into distinct dynamical regimes. On naturalistic sensory streams, the online algorithm produces center–surround spatial receptive fields and ON/OFF sustained and transient temporal filters resembling those of early visual neurons. Our results suggest that predictive learning can provide a self-supervised mechanism by which neural circuits discover the local geometry of environmental dynamics and represent it in receptive fields.
\end{abstract}


\section{Introduction}\label{sec:intro}

High-dimensional time series from physical, biological, and neural systems often evolve on structured nonlinear manifolds. Global dynamics may be complex, yet short-time evolution is typically low-dimensional: within a small neighborhood, state changes are captured by a few tangent directions --- the local-linear view underlying classical nonlinear dynamical systems theory \citep{strogatz1994nonlinear}, Koopman and dynamic mode decomposition \citep{schmid2010dynamic,brunton2021modern}, and local models of nonlinear time series \citep{farmer1987predicting,casdagli1991chaos}. Some tangent directions are more predictive than others: perturbations along them set the system's evolution over the next few steps, while orthogonal ones decay, fluctuate, or say little about the future. This suggests a self-supervised problem: from an unlabeled trajectory alone, can one discover tangent directions that vary slowly along it and whose projections predict future evolution?\looseness=-1

Near a saddle point, for instance, unstable manifolds organize future trajectories: projecting the state, relative to the saddle, onto an unstable direction determines which branch is followed \citep{strogatz1994nonlinear}. Such saddle-organized dynamics are central to stable heteroclinic channels, winnerless competition, and transient neural dynamics \citep{rabinovich2001dynamical,rabinovich2006dynamical,rabinovich2008transient,cvitanovic2012knowing}. Away from fixed points, short trajectory segments still sample a local linearization, so the predictive objects are not global coordinates but local tangent-space directions varying across state space.\looseness=-1

Existing approaches fall into three families (details in Appendix~\ref{app:related}). Switching linear dynamical systems and spectral HMM methods \citep{ackerson1970state,fox2009nonparametric,linderman2017bayesian,hsu2012spectral} require explicit inference over latent discrete states under a generative model. Koopman operator theory, dynamic mode decomposition and VAMP \citep{brunton2021modern,schmid2010dynamic,wu2020variational}, and their deep variants \citep{lusch2018deep,otto2019linearly,takeishi2017learning}, seek a globally linearizing feature space whose leading modes span broad regions of state space and may not resolve local saddle-to-saddle geometry. Deep predictive objectives \citep{mardt2018vampnets,oord2018representation,zerveas2020mask} learn global embeddings, but are heavily parameterized and less interpretable.\looseness=-1

\begin{figure}[t]
\centering
\includegraphics[width=1\columnwidth]{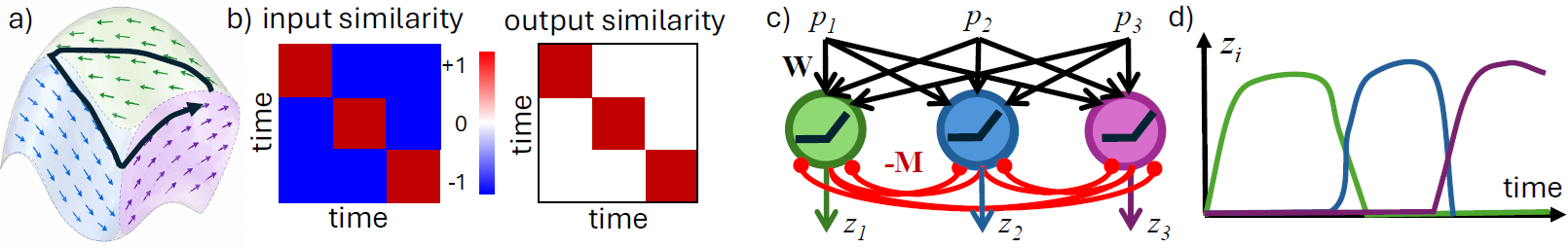}
\setlength{\abovecaptionskip}{-2mm}
\setlength{\belowcaptionskip}{-6mm}
    \caption{\textbf{Overview of the LPD algorithm.}
(a) Clusters of slowly varying flow directions in the tangent space of a nonlinear dynamical system (colors), with a sample trajectory in black.
(b) Similarity of temporal differences (left) and of the non-negative outputs (right) across time.
(c) Online implementation as a self-supervised neural network, with feedforward weights $\mathbf{W}$ and lateral weights $\mathbf{M}$ (see \emph{Online Algorithm and Neural Interpretation}, Section~\ref{sec:lpd}).
(d) Soft domain assignments over time for the trajectory in (a).}
    \label{saddles}
\end{figure}

Past--future correlation offers a principled definition of such predictive directions. Canonical correlation analysis (CCA) finds projections of paired variables with maximal correlation \citep{hotelling1936relations}; related past--future covariance methods underlie subspace system identification and balanced stochastic realization \citep{arun1990balanced,katayama2005subspace}, and connect to predictive information and the Gaussian information bottleneck \citep{bialek2001predictability,chechik2005information}. Applied to past and future observations, it finds projections of the present maximally informative about subsequent evolution, recovering dominant predictive modes in locally linear dynamics. This motivates the Rectified Spectral Unit (ReSU), a biologically inspired self-supervised rule combining past--future CCA with rectification to learn predictive features from dynamical data \citep{pughe2026neurons,Qin2026}. Existing CCA- and ReSU-based methods are nonetheless global or single-regime: they assume one predictive structure dominates, or that local neighborhoods are supplied externally, and so do not discover multiple local predictive directions whose locations and domains are unknown.\looseness=-1

We introduce a self-supervised objective and algorithm for learning such local predictive directions (LPDs) directly from trajectories, Fig. \ref{saddles}. Temporal differences give access to local tangent-space information: applying past--future CCA to finite differences rather than absolute states makes the method invariant to translations of the local coordinate system, such as a saddle's unknown location, so predictive directions are learned from trajectory increments without estimating equilibria or fitting a global dynamical model. To discover multiple local predictive regimes, we replace canonical projection followed by rectification with a non-negative similarity-matching formulation. Similarity matching provides a normative route from similarity-preserving objectives to online neural learning algorithms \citep{pehlevan2015hebbian,pehlevan2019neuroscience,pehlevan2020neurons,lipshutz2021biologically}, while non-negativity promotes localized, parts-based, and clustering-like representations \citep{lee1999learning,pehlevan2014hebbian,sengupta2018learning}. Here, similarity matching preserves the predictive correlation structure of the data while non-negativity induces competition and locality, Fig. \ref{saddles}b; output units therefore specialize to different regions of state space, learning distinct local predictive directions, Fig. \ref{saddles}c, and their non-negative activities act as soft domain assignments, Fig. \ref{saddles}d. Unlike clustering by geometric proximity or variance \citep{reynolds2009gaussian}, the proposed objective clusters trajectory segments by short-time predictive structure. The resulting algorithm admits offline and online implementations, the online one having a natural neural interpretation in terms of feedforward weights, lateral competition, and rectified activity.\looseness=-1

We validate LPD on stochastic nonlinear systems with known local structure and on sensory streams from translated models of natural images, where the learned temporal filters resemble those of biological neurons and pooling across pixels yields localized filters that tile space like biological receptive fields. These results extend previous predictive interpretations of neural computation~\citep{srinivasan1982predictive,rao1999predictive,chalk2018toward,rao2024sensory}.\looseness=-1

\section{Learning Local Predictive Directions from Trajectory Data}\label{sec:lpd}
\paragraph{Local Predictive Structure in Tangent Space.}
Our goal is to identify local directions in state space whose projections are maximally informative about near-future evolution of a stochastic discrete-time dynamical system, $\mathbf{x}_{t+\zeta} = \mathbf{f}(\mathbf{x}_t) + \bm{\omega}_t$, where $\mathbf{x}_t \in \mathbb{R}^n$ and $\bm{\omega}_t$ denotes process noise.
Over a sufficiently short trajectory segment, the dynamics can be approximated locally by an affine map. Around a reference point $\mathbf{x}_*$ in state space,
\begin{equation}
\label{eq:local_affine}
    \mathbf{x}_{t+\zeta} - \mathbf{f}(\mathbf{x}_*)
    \approx
    \mathbf{A}_{\mathbf{x}_*}(\mathbf{x}_t-\mathbf{x}_*) + \bm{\omega}_t ,
    \qquad
    \mathbf{A}_{\mathbf{x}_*}
    :=
    \left. \frac{\partial \mathbf{f}}{\partial \mathbf{x}} \right|_{\mathbf{x}_*}.
\end{equation}
Locally predictive directions in tangent space determine the subsequent displacement of the trajectory and can be viewed as low-rank components of the local linear map $\mathbf{A}_{\mathbf{x}_*}$. In nonlinear systems, these directions generically vary across state space. A trajectory may pass through multiple regions in which different tangent directions are predictive (Fig.~\ref{saddles}a). The learning problem is therefore twofold: to discover a finite set of local predictive directions, and to assign trajectory segments to the regions in which each direction is active. A direct approach based on Eq.~\eqref{eq:local_affine}, however, requires representing states relative to the appropriate local reference point $\mathbf{x}_*$, which is generally unknown from data.\looseness=-1

\paragraph{Translation-Invariant Predictive Learning from Temporal Differences.}
To remove the dependence of Eq.~\eqref{eq:local_affine} on unknown local reference points, we represent local dynamics using temporal differences. For a fixed lag $\tau$, define $\delta \mathbf{x}_t := \mathbf{x}_{t+\tau} - \mathbf{x}_t$. If the dynamics are approximately affine over the corresponding local trajectory segment, then
\begin{equation}
\begin{split}
    \delta \mathbf{x}_{t+\zeta}
    =
    \mathbf{x}_{t+\zeta+\tau} - \mathbf{x}_{t+\zeta}
    \approx
    \mathbf{A}_{\bf{x}_*}(\mathbf{x}_{t+\tau}-\mathbf{x}_t)
    +
    (\bm{\omega}_{t+\tau}-\bm{\omega}_t)
    =
    \mathbf{A}_{\bf{x}_*}\delta \mathbf{x}_t + \bm{\eta}_t ,
\end{split}
\end{equation}
where $\bm{\eta}_t:=\bm{\omega}_{t+\tau}-\bm{\omega}_t$ is an effective differenced-noise term. Importantly, the affine offset cancels. Thus, temporal differences preserve the local linear action of $\mathbf{A}_{\bf{x}_*}$ on tangent displacements while removing the dependence on unknown translations of the local coordinate system. This inspires us to define past and future descriptors from tangent space rather than absolute
states, $\mathbf{p}_t := \delta \mathbf{x}_t$ and $\mathbf{f}_t := \delta \mathbf{x}_{t+\zeta}$,
which is translation invariant, and in fact invariant to any invertible affine change of coordinates (Appendix~\ref{app:invariance}).\looseness=-1

\paragraph{Predictive Directions via Past--Future Correlation.}
Global predictive directions can be found via past--future canonical correlation analysis (CCA)~\citep{arun1990balanced,chechik2005information}. Here, to find local predictive directions we apply CCA to the temporal differences. Specifically, we seek directions $\mathbf{v}_i$ and $\mathbf{u}_i$ that maximize the correlation between linear projections of $\mathbf{p}_t$ and $\mathbf{f}_t$:
\begin{equation}
    \mathbf{u}_1, \mathbf{v}_1
    =
    \arg\max_{\mathbf{u},\mathbf{v}}
    \mathbb{E}\!\left[(\mathbf{v}^\top \mathbf{p}_t)(\mathbf{u}^\top \mathbf{f}_t)\right]
    \quad
    \text{s.t.}
    \quad
    \mathbb{E}\!\left[(\mathbf{v}^\top \mathbf{p}_t)^2\right]
    =
    \mathbb{E}\!\left[(\mathbf{u}^\top \mathbf{f}_t)^2\right]
    =
    1 ,
\end{equation}
with additional orthogonality constraints for subsequent components. Define the covariance matrices $\mathbf{C}_{pp}:=\mathbb{E}[\mathbf{p}_t \mathbf{p}_t^\top]$, $\mathbf{C}_{ff}:=\mathbb{E}[\mathbf{f}_t \mathbf{f}_t^\top]$, $\mathbf{C}_{pf}:=\mathbb{E}[\mathbf{p}_t \mathbf{f}_t^\top]$ and $\mathbf{C}_{fp}:=\mathbb{E}[\mathbf{f}_t \mathbf{p}_t^\top]$. The CCA directions solve
\begin{equation} \label{eq:cca_eigenproblem}
    \mathbf{C}_{pp}^{-1} \mathbf{C}_{pf} \mathbf{C}_{ff}^{-1} \mathbf{C}_{fp} \mathbf{v}_i
    =
    \rho_i^2 \mathbf{v}_i ,
    \qquad
    \mathbf{C}_{ff}^{-1} \mathbf{C}_{fp} \mathbf{C}_{pp}^{-1} \mathbf{C}_{pf} \mathbf{u}_i
    =
    \rho_i^2 \mathbf{u}_i ,
\end{equation}
where $\rho_i$ is the $i$th canonical correlation coefficient.
When the local dynamics are dominated by a single expanding or otherwise predictive mode, the leading canonical direction captures the strongest evolving tangent-space direction (Appendix \ref{app:saddle_case}). Near a saddle dominated by one unstable mode, the leading past filter extracts the unstable coordinate (the corresponding left eigenvector), while the associated future mode aligns with the unstable tangent direction~\citep{pughe2026neurons}.\looseness=-1

\paragraph{CCA as a Similarity-Matching Objective.}
To construct an online algorithm with a natural neural interpretation, we now turn the predictive-correlation principle into a self-supervised learning objective. Let $\mathbf{P} := [\mathbf{p}_1,\ldots,\mathbf{p}_T] \in \mathbb{R}^{n \times T}$ and $\mathbf{F} := [\mathbf{f}_1,\ldots,\mathbf{f}_T] \in \mathbb{R}^{n \times T}$ denote matrices of past and future increment descriptors. For the top $k$ canonical components, define $\mathbf{Z}_p := \mathbf{V}^\top \mathbf{P}$ and $\mathbf{Z}_f := \mathbf{U}^\top \mathbf{F}$, where $\mathbf{V}=[\mathbf{v}_1,\ldots,\mathbf{v}_k]$ and $\mathbf{U}=[\mathbf{u}_1,\ldots,\mathbf{u}_k]$ contain the CCA directions. Define the predictive similarity matrix
\begin{equation}
    \mathbf{G}
    :=
    \mathbf{P}^\top \mathbf{C}_{pp}^{-1} \mathbf{C}_{pf} \mathbf{C}_{ff}^{-1} \mathbf{F}
    = \mathbf{Z}_p^\top \mathbf{R}\mathbf{Z}_f ,
\end{equation}
where $\mathbf R_{ij}=\rho_i\delta_{ij}$ is the diagonal matrix of canonical correlations and the
second equality is derived in Appendix \ref{app:normalization} for the full-rank case $k=n$,
becoming a rank-$k$ approximation for $k<n$. This matrix measures how strongly each past sample predicts each future sample after whitening the past and future representations (Appendix \ref{app:whitened}). $\mathbf{G}$ requires only sample covariances of the increments; the Jacobian $\mathbf{A}_{\mathbf{x}_*}$ is never estimated. For the unconstrained objective~\eqref{eq:unconstrained_objective} below, the leading canonical coordinates span the dominant singular subspace of this predictive similarity matrix (Appendix \ref{app:subspace}). Thus, rather than solving the CCA eigenproblem explicitly, we can recover the predictive representation by solving the following similarity-matching problem over trajectory segments:
\begin{equation}
    \label{eq:unconstrained_objective}
    \mathbf{Z}^*
    =
    \arg\min_{\mathbf{Z}\in\mathbb{R}^{k\times T}}
    \left\|
        \mathbf{G}\mathbf{G}^\top - \mathbf{Z}^\top \mathbf{Z}
    \right\|_F^2 .
\end{equation}
The minimizer $\mathbf{Z}^*$ spans the same canonical subspace as $\mathbf{Z}_p$, up to an orthogonal transformation; its component scaling is set by the spectrum of $\mathbf{G}\mathbf{G}^\top$
 and differs from the conventional CCA normalization. Intuitively, two time points should have similar output representations if their past increments predict future increments in similar ways. This similarity matching formulation of CCA is different from previous ones~\citep{pehlevan2020neurons,lipshutz2021biologically} in that it allows a causal online implementation (Algorithm~\ref{online_CCA}) that computes the output from the past input alone while using future input for learning only.

\paragraph{Learning Multiple Local Predictive Directions with Non-Negative Similarity Matching.}
The preceding objective \eqref{eq:unconstrained_objective} learns a global predictive subspace. However, nonlinear dynamics typically contain multiple local predictive regimes: different regions of state space can have different tangent directions that predict future motion (Fig. \ref{saddles}a). A single unconstrained CCA subspace can therefore mix together distinct local dynamical structures. To separate these regimes, we impose elementwise non-negativity on the output representation:
\begin{equation}
\label{eq:non_neg}
    \min_{\mathbf{Z} \geq 0}
    \left\|
        \mathbf{G}\mathbf{G}^\top - \mathbf{Z}^\top \mathbf{Z}
    \right\|_F^2 .
\end{equation}
Non-negativity changes the interpretation of the output~\citep{pehlevan2014hebbian,pehlevan2017clustering,sengupta2018learning}. Instead of signed canonical coordinates, the columns of $\mathbf{Z}$ become soft domain assignments: $\{\mathbf{Z}\}_{it}$ indicates the degree to which trajectory segment $t$ belongs to predictive regime $i$ (Fig. \ref{saddles}d). The similarity-matching term strives to preserve past--future predictive structure, while non-negativity and lateral competition encourage different units to specialize to different local regions of state space. Unlike clustering in state space~\citep{pehlevan2014hebbian,pehlevan2017clustering,sengupta2018learning}, it groups trajectory segments whose past increments predict their future increments in similar ways, discovering domains with shared local predictive geometry.
Applying the Legendre transform, we derive from Eq. \eqref{eq:non_neg} the following equivalent min--max problem over auxiliary variables $\mathbf{N}\in\mathbb{R}^{k\times T}$ and $\mathbf{M}\in\mathbb{R}^{k\times k}$, whose optima are $\mathbf{N}^*=\mathbf{Z}\mathbf{G}$ and $\mathbf{M}^*=\mathbf{Z}\mathbf{Z}^\top$ (Appendix \ref{app:legendre}).
\begin{equation}
\label{eq:min-max legend}
\begin{split}
    \min_{\mathbf{Z}\geq 0} \min_{\mathbf{N}} \max_{\mathbf{M}}
    \operatorname{Tr}
    \Big[
        -4 \mathbf{Z}^\top \mathbf{N} \mathbf{G}^\top
        + 2 \mathbf{Z}^\top \mathbf{M} \mathbf{Z}
        + 2 \mathbf{N}^\top \mathbf{N}
        - \mathbf{M}^\top \mathbf{M}
    \Big].
\end{split}
\end{equation}
Optimization of Eq.~\eqref{eq:min-max legend} using gradient descent alternates an inference step for $\mathbf{Z}$ at fixed $\mathbf{W},\mathbf{M}$ with updates of $\mathbf{W}$ and $\mathbf{M}$ from empirical covariances, where the weight matrix $\mathbf{W}$ is defined by $\mathbf{W}\mathbf{P} = \mathbf{N}\mathbf{G}^\top$; the resulting batch algorithm is given in Appendix~\ref{app:numerics}.\looseness=-1

\paragraph{Online Algorithm and Neural Interpretation.} Since all covariances can be updated recursively, objective~\eqref{eq:non_neg} also admits an online algorithm: at each step it receives $(\mathbf{p}_t,\mathbf{f}_t)$, infers a non-negative activity $\mathbf{z}_t$, and updates the covariances, $\mathbf{W}$ and $\mathbf{M}$ (Algorithm~\ref{online_CCA}).
\begin{algorithm2e}[t]
\caption{Learning Local Predictive Directions Online. $\mathbf{C}^{+}:=(\mathbf{C}+\epsilon\mathbf{I})^{-1}$ denotes a Tikhonov-regularized inverse and $[\cdot]_+$ elementwise rectification.}
\label{online_CCA}
Initialize $\mathbf{W} \in \mathbb{R}^{k \times n}$ and $\mathbf{M} \in \mathbb{R}^{k \times k}$ with $\mathbf{M} \succ 0$\;
Initialize covariance estimates
$\mathbf{C}_{pp},\mathbf{C}_{ff},\mathbf{C}_{pf},\mathbf{C}_{fp},\mathbf{C}_{zp},\mathbf{C}_{zz}$\;
\For{$t\leftarrow 1$ \KwTo $T$}{
  Receive increment descriptors $(\mathbf{p}_t,\mathbf{f}_t)$\;
  Update $\mathbf{C}_{pp},\mathbf{C}_{ff},\mathbf{C}_{pf},\mathbf{C}_{fp}$\;
  \textbf{Inference:} update $\mathbf{z}_t$ for fixed $\mathbf{W},\mathbf{M}$\;
  \Repeat{inference convergence}{
    $\mathbf{z}_t \leftarrow [\mathbf{z}_t + \eta_Z(\mathbf{W}\mathbf{p}_t - \mathbf{M}\mathbf{z}_t)]_+$\;
  }
  Update $\mathbf{C}_{zp}$ and $\mathbf{C}_{zz}$\;
  Update $\mathbf{W} \leftarrow \mathbf{C}_{zp} \mathbf{C}_{pp}^{+} \mathbf{C}_{pf}
  \mathbf{C}_{ff}^{+} \mathbf{C}_{fp} \mathbf{C}_{pp}^{+}$\;
  Update $\mathbf{M} \leftarrow \mathbf{C}_{zz}$\;
}
\end{algorithm2e}
This online algorithm has a natural neural interpretation (Fig. \ref{saddles}c). The matrix $\mathbf{W}$ corresponds to feedforward synaptic weights that define local predictive filters over past increments. The matrix $\mathbf{M}$ corresponds to lateral interactions that implement competition between units. The circuit interpretation applies to the inference dynamics; the weight updates, which are non-local, provide an efficient streaming implementation. Rectification enforces non-negative activity, so each unit reports the presence of a particular local predictive direction rather than a signed canonical coordinate. The resulting activity vector $\mathbf{z}_t$ therefore acts as a soft domain assignment of the current trajectory segment to one of several locally predictive tangent-space regimes (Fig. \ref{saddles}d).\looseness=-1

\section{Experiments on Stochastic Dynamical Systems}\label{sec:dynamical_systems_exp}
\begin{figure}[t]
    \centering
\includegraphics[width=1\linewidth]{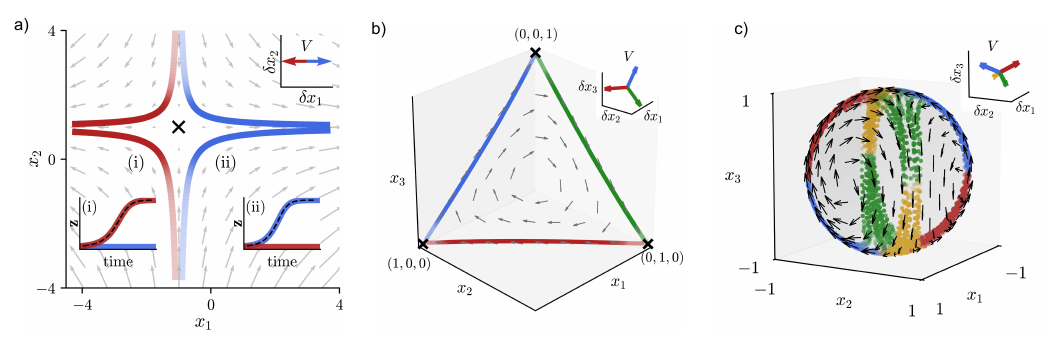}
\setlength{\abovecaptionskip}{-3mm}
\setlength{\belowcaptionskip}{-2mm}
\caption{\textbf{Online LPD (Alg.~\ref{online_CCA}) applied to stochastic dynamical systems.}
Gray arrows show the flow; trajectories are colored by the most active output, saturation by its magnitude; insets show the learned tangent-space projections.
(a) \textit{Affine dynamics near a saddle at $(-1,1)$}, $k=2$; insets (i)--(ii) compare the output with the expected activation (black dashed) for two trajectories.
(b) \textit{gLV} with $N=3$ species, $k=3$. (c) \textit{Two-saddle dynamics on a sphere}, $k=4$.}
    \label{fig:dynamical_systems}
\end{figure}

\paragraph{Affine saddle-point dynamics.}
For affine dynamics near a single saddle point, Fig.~\ref{fig:dynamical_systems}a,
ReSU~\citep{pughe2026neurons} fails to recover the unstable directions unless the data are provided relative to the fixed-point location. In contrast, our algorithm correctly identifies the unstable directions even when the fixed-point location is unknown (see Appendices~\ref{app:saddle_case} and~\ref{app:affine}).
The output activity, $\mathbf{z}_t \in \mathbb{R}_+^{2}$, represents the rectified projection of the past finite differences $\mathbf{p}_t \in \mathbb{R}^{2}$ onto the unstable manifold.

\paragraph{Generalized Lotka--Volterra Systems.}
To test multiple saddle points, we apply LPD to a stochastic generalized Lotka--Volterra (gLV) system, widely used to study stable heteroclinic channels (SHC)~\citep{bick2010occurrence,afraimovich2008winnerless}. We consider $N=3$ species arranged to form an SHC, producing sequential transitions between saddles at which one species dominates and the others are nearly extinct (Appendix~\ref{app:dynamical_systems}). The resulting dynamics generate a limit-cycle-like sequence of metastable states, as shown in Fig.~\ref{fig:dynamical_systems}b. The results of Algorithm~\ref{online_CCA} show that each of the $k=3$ output units successfully captures the predictive direction associated with one transition between saddles, thereby predicting the identity of the next dominant species. These findings remain robust for a greater number of species ($N\le 100$) and across different noise levels (see Appendix~\ref{app:high_dimensional_glv}).\looseness=-1

\paragraph{Two Saddle Points on a Sphere.}\label{sect:sphere_example}
On a curved manifold, we consider a heteroclinic flow on the sphere~\citep{peixoto1962structural, batchelor2000introduction} with saddle points at the north and south poles (Appendix~\ref{app:dynamical_systems}). LPD recovers the predictive geometry without prior knowledge of the saddle locations, Fig.~\ref{fig:dynamical_systems}c. The symmetry of this system also reveals a limitation: the unstable direction at one pole aligns with the stable direction in the antipodal region, leading to redundant activations. Such aligned regimes cannot be distinguished from tangent information alone and require additional state or temporal context.\looseness=-1

\paragraph{Model components and baselines.}
The objective \eqref{eq:non_neg} itself provides an internal criterion for the number of output
components $k$, unlike the external criteria of $k$-means clustering or Hidden Markov Models: the objective plateaus once $k$ reaches the intrinsic dimensionality of the predictive structure, and this heuristic elbow coincides with the precision and recall optima of the learned directions (Appendix~\ref{app:model_selection}; Appendix~\ref{app:modelsel_stimuli} for natural stimuli). To isolate the roles of temporal differencing and
non-negativity, we compare LPD on the gLV system (Fig.~\ref{fig:dynamical_systems}b) against
past--future CCA (PF), non-negative PF (NN-PF) and finite-differences CCA (FD), scoring held-out
trajectories by the Normalized Mutual Information (NMI) between the maximally active component
and the next saddle visited. At low and intermediate noise LPD attains higher NMI than all three,
converging with FD only at high noise, so both ingredients matter (Appendix~\ref{app:ablation}).
LPD also carries more information about the future tangent vector than a Gaussian Mixture Model
fit to the same finite differences, partly via its unnormalized magnitude
(Appendix~\ref{app:clustering}).\looseness=-1

\section{Learning Neuronal Responses from Natural Stimuli}\label{sec:natural_stimuli}
\begin{figure}[t]
    \centering    \includegraphics[width=1\linewidth]{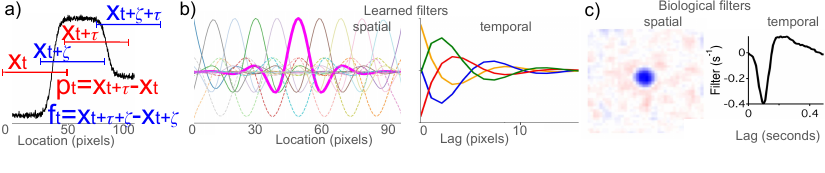}
    \setlength{\abovecaptionskip}{-4mm}
    \setlength{\belowcaptionskip}{-2mm}
   \caption{\textbf{LPDs learned on naturalistic stimuli resemble biological receptive fields and temporal filters.} (a) Translated ``dead leaves'' stimulus; past and future samples are temporal differences between segments.
    (b) Left: center--surround ON (solid) and OFF (dashed) LPDs tile space ($k=20,\,\tau=1,\,\zeta=2$). Right: LPDs on lag vectors ($k=4,\,\tau=1,\,\zeta=n+1$, $n$ the embedding dimension) resemble sustained ON/OFF (green/blue) and transient ON/OFF (red/yellow) cells.
    (c) Measured receptive field and temporal filter of a retinal bipolar cell (courtesy of D.B. Kastner and S.A. Baccus).}
\label{fig:receptive_fields}
\end{figure}

Here, we apply Algorithm~\ref{online_CCA} to sensory stimuli, viewing them as observations of an unknown nonlinear dynamical system. Over short timescales, much of the variation perceived by the retina arises from self-motion and saccadic eye movements. This motivates modeling natural scenes as time series formed by sliding windows over static images.
For simplicity, we use the one-dimensional ``dead leaves'' model of natural images~\citep{jeulin1996dead,ruderman1997origins}, which partitions space into patches of different but uniform contrast with sharp boundaries (Fig.~\ref{fig:receptive_fields}a).
First, we apply LPD to the past and future temporal differences of pairs of translated images assuming they represent dynamical states, Fig.~\ref{fig:receptive_fields}a. The LPDs qualitatively resemble electrophysiologically measured neuronal receptive fields: i) The non-negative constraint along with the short-range correlations in natural images drives each filter to be localized and tile the pixel space (Fig. \ref{fig:receptive_fields}b, left), which corresponds to the retinal mosaic \citep{werblin1969organization, kastner2011coordinated}. ii) The center--surround structure of each filter in Fig. \ref{fig:receptive_fields}b, left resembles the spatial receptive field of bipolar cells~\citep{hartline1938response, kuffler1953discharge, barlow1953summation}, Fig.~\ref{fig:receptive_fields}c, left. iii) Filters with positive and negative peaks correspond to ON and OFF bipolar cells. Each class tiles the pixel space independently of the other~\citep{kolb1976rod}. Unlike previous explanations of the ON/OFF center--surround receptive fields based on efficient coding~\citep{karklin2011efficient,jun2022efficient,gjorgjieva2019functional}, our approach does not rely on explicitly minimizing metabolic costs.
Second, we apply LPD to the past--future temporal difference pairs of time-delay embedding of a scalar time series obtained from the same ``dead leaves'' model. Such time-delay embedding is necessary if the dynamical system is only partially observed~\citep{takens1981detecting}.
Within this framework, the learned weights define linear temporal filters mapping past finite-difference patterns to future ones. Resulting filters (Fig.~\ref{fig:receptive_fields}b, right) qualitatively resemble ON/OFF sustained/transient filters of retinal bipolar cells (Fig.~\ref{fig:receptive_fields}c, right)~\citep{kuffler1953discharge, werblin1969organization,cleland1973properties}.
Both filter types can be interpreted as LPDs of the stimulus dynamics, which alternate between two local regimes: plateaus of nearly constant luminance, and transitions where contrast boundaries move across the receptive field. Plateaus yield LPDs that integrate the input over time (sustained filters); transitions yield LPDs that compute temporal derivatives (transient filters) and spatial derivatives (center--surround receptive fields), reflecting the inertia of moving boundaries: the recent direction of motion predicts the near future. Spatial and temporal filters differ because motion directions are nearly isotropic in space, whereas time has a fixed arrow. These results suggest that sensory neurons can be viewed as detectors of LPDs governing stimulus dynamics.\looseness=-1

\section{Discussion}
We introduced a data-driven framework for discovering LPDs in dynamical systems. Across synthetic and naturalistic trajectory data, this approach identifies LPDs without explicit system identification or latent-state inference, providing a simple primitive for identifying predictive structure (LPD does not itself produce forecasts).  Because learning relies only on second-order short-time statistics and a compact set of feedforward and lateral weights, LPD is substantially less parameterized than typical deep predictive architectures, an advantage when data are limited.
The current framework assumes that dynamics are approximately linear within local neighborhoods and that predictive directions associated with distinct regimes are sufficiently separated to be resolved by competition. Systems with extremely fast switching, strong nonlinearities, or heavy-tailed noise may violate these assumptions and require extensions of the objective. An important direction for future work is to establish theoretical conditions under which non-negative similarity matching provably recovers LPDs.
Finally, the learned LPDs correspond to modes where small perturbations influence future trajectories, suggesting a natural interface with control. Low-rank controllers acting along these directions may enable efficient steering between metastable regions, analogous to control along unstable manifolds in classical dynamical systems. Incorporating reward-weighted or goal-conditioned variants of the predictive objective could therefore yield a self-supervised route to learning control-relevant representations without explicit system identification. Our framework may thus help model neural computation beyond the sensory periphery.\looseness=-1

\bibliography{example_paper}
\clearpage
\appendix

\paragraph{Code availability.} Code to reproduce all experiments is available at \url{https://github.com/MarcoZenari/local-predictive-directions}.

\section{Extended Related Work}\label{app:related}

Our method is related to existing approaches for modeling dynamical systems, but differs in key aspects. Switching Linear Dynamical Systems (SLDS) \citep{ackerson1970state,fox2009nonparametric,linderman2017bayesian} model nonlinear dynamics as collections of locally linear regimes indexed by a latent discrete state that evolves according to a Markov chain.  \citet{hsu2012spectral} use an SVD of past–future observation probabilities to recover a global observable representation of a Hidden Markov Model (HMM). Whereas these approaches require explicit inference over latent states and optimize the likelihood of observations under a generative model, we discover LPDs directly from short-time correlations and output soft domain assignments online.

An alternative to explicitly partitioning state space is to approximate nonlinear dynamics with linear dynamics in a transformed feature space. Koopman operator theory \citep{brunton2021modern}, Dynamic Mode Decomposition (DMD) \citep{schmid2010dynamic}, and the Variational Approach for Markov Processes (VAMP) \citep{wu2020variational} all pursue this strategy, aiming to represent nonlinear dynamics with global modes of a linear operator acting on a diverse, and potentially high dimensional, set of observables. Classic finite-rank approaches, such as DMD and VAMP, are powerful and interpretable, but often emphasize leading eigen- or singular modes, which in many applications correspond to slow, large-scale, or long-timescale structure. As a result, the learned dynamic modes can span broad regions of state space and may not resolve the local saddle-to-saddle geometry that governs transitions in multistable systems. Recent deep Koopman methods go beyond fixed dictionaries and leading spectral truncations by learning expressive nonlinear observables or encoder–decoder representations in which the dynamics are approximately linear \citep{lusch2018deep, otto2019linearly, takeishi2017learning}. However, these methods still primarily seek a globally valid transformed state-space representation. Most relevant to our setting, \citet{lusch2018deep} learn state-dependent latent eigenvalues through an auxiliary network, enabling local variation in the learned Koopman dynamics. This construction is especially natural for low-dimensional oscillatory or continuous-spectrum systems, where latent dynamics can be represented by structured rotating or spiraling modes, but may be less directly suited to transient dynamics that move between unstable centers. In contrast, LPD is local in both space and time: by operating on finite differences, it directly decomposes trajectory segments into localized predictive directions and non-negative soft domain assignments, readily accommodating multiple unstable centers.

In addition, deep learning methods have extended prior operator-based and variational frameworks (e.g., VAMPnets \citep{mardt2018vampnets}) and have been applied to multivariate time series forecasting via Transformer architectures. Representative Transformer-style objectives include contrastive predictive coding (CPC) \citep{oord2018representation} and masked modeling \citep{zerveas2020mask} which learn global predictive embeddings. These models are highly parameterized and trained on large datasets with task-specific training procedures, often requiring significant computational costs and careful hyperparameter tuning. While these models can achieve strong predictive performance, the resulting representations are typically global and less interpretable, encompassing complex dependencies across the input sequence. In contrast to global latent-space models and Gaussian Mixture Models \citep{reynolds2009gaussian}, the self-supervised method proposed here identifies LPDs from the system’s short time correlations.

\section{CCA Recovers the Most Predictive Direction of a Saddle Fixed Point} \label{app:saddle_case}
We start from Eq. \eqref{eq:cca_eigenproblem} in the main text:
\begin{align}
\label{sigma}
\mathbf{C}_{pp}^{-1}\mathbf{C}_{pf}\mathbf{C}_{ff}^{-1}\mathbf{C}_{fp}\mathbf{v}_i = \rho_i^2\,\mathbf{v}_i,\quad
\mathbf{C}_{ff}^{-1}\mathbf{C}_{fp}\mathbf{C}_{pp}^{-1}\mathbf{C}_{pf}\mathbf{u}_i = \rho_i^2\,\mathbf{u}_i.
\end{align}
Under the linearized dynamics,
\begin{equation}
\begin{aligned}
\mathbf{C}_{pf} &= \mathbb{E}[\mathbf{p}_t(\mathbf{A}\mathbf{p}_t+\boldsymbol{\omega}_t)^\top]
= \mathbf{C}_{pp}\mathbf{A}^\top,\\
\mathbf{C}_{fp} &= \mathbb{E}[(\mathbf{A}\mathbf{p}_t+\boldsymbol{\omega}_t)\mathbf{p}_t^\top]
= \mathbf{A}\mathbf{C}_{pp},
\end{aligned}
\end{equation}
which reduces Eq. \eqref{eq:cca_eigenproblem} to
\begin{equation}
\label{reduced_cca}
\begin{aligned}
\mathbf{A}^\top\mathbf{C}_{ff}^{-1}\mathbf{A}\mathbf{C}_{pp}\mathbf{v}_i &= \rho_i^2\,\mathbf{v}_i,\\
\mathbf{C}_{ff}^{-1}\mathbf{A}\mathbf{C}_{pp}\mathbf{A}^\top \mathbf{u}_i &= \rho_i^2\,\mathbf{u}_i.
\end{aligned}
\end{equation}
Notice that
\begin{equation}
\begin{aligned}
    &(\mathbf{A}^\top\mathbf{C}_{ff}^{-1}\mathbf{A}\mathbf{C}_{pp})^\top \mathbf{C}_{pp} \mathbf v_i\\
    =&\mathbf{C}_{pp} \mathbf{A}^\top\mathbf{C}_{ff}^{-1} \mathbf{A}\mathbf{C}_{pp}\mathbf v_i\\
    =&\rho_i^2\mathbf{C}_{pp}\mathbf v_i.
\end{aligned}
\end{equation}
Thus, $\{\mathbf v_i, \mathbf{C}_{pp}\mathbf v_i\}$ is a pair of right and left eigenvectors of $\mathbf{A}^\top\mathbf{C}_{ff}^{-1}\mathbf{A}\mathbf{C}_{pp}$. Similarly, $\{\mathbf u_i, \mathbf{C}_{ff}\mathbf u_i\}$ is a pair of right and left eigenvectors of $\mathbf{C}_{ff}^{-1}\mathbf{A}\mathbf{C}_{pp}\mathbf{A}^\top$. Therefore, $\mathbf u_i^\top \mathbf{C}_{ff}\mathbf u_j=\mathbf v_i^\top\mathbf{C}_{pp}\mathbf v_j=0 \text{ for } i\neq j$.\\
Also notice that
\begin{equation}
\begin{aligned}
    &\mathbf{A}^\top(\mathbf{C}_{ff}^{-1}\mathbf{A}\mathbf{C}_{pp}\mathbf{A}^\top \mathbf u_i)\\
    =&(\mathbf{A}^\top\mathbf{C}_{ff}^{-1}\mathbf{A}\mathbf{C}_{pp})\mathbf{A}^\top \mathbf u_i\\
    =&\rho_i^2 \mathbf{A}^\top\mathbf u_i.
\end{aligned}
\end{equation}
Thus, $\mathbf v_i\propto \mathbf{A}^\top\mathbf u_i$. Under proper normalization, these results can be summarized by
\begin{align}
    \mathbf{V}^\top\mathbf{C}_{pp} \mathbf{V}=\mathbf{I},\quad \mathbf{U}^\top\mathbf{C}_{ff} \mathbf{U}=\mathbf{I},\quad \mathbf{A}^\top \mathbf{U}= \mathbf{V}\mathbf{R},
\end{align}
where $\mathbf{U}[:,i]\coloneq \mathbf{ u}_i$, $\mathbf{V}[:,i]\coloneq \mathbf v_i$, $\mathbf{R}_{ij}\coloneq \delta_{ij}\rho_i$. Thus, we have
\begin{align}
    &\mathbf{A}^\top=\mathbf{V}\mathbf{R} \mathbf{U}^{-1}=\mathbf{V}\mathbf{R} \mathbf{U}^\top \mathbf{C}_{ff},\\
    &\mathbf{A}=\mathbf{C}_{ff} \mathbf{U} \mathbf{R} \mathbf{V}^\top=\sum_i \rho_i\mathbf{C}_{ff} \mathbf{u}_i\mathbf{v}_i^\top.
    \label{svd_of_e^(At)}
\end{align}
On the other hand, we have the eigen-decomposition
\begin{align}
    \mathbf{A}=\sum_i \lambda_i\mathbf r_i \mathbf l_i^\top,
    \label{eigen_decomposition_of_e^(At)}
\end{align}
where $\lambda_i$ is the eigenvalue of $\mathbf{A}$, $\mathbf r_i$ and $\mathbf l_i$ are right and left eigenvectors of $\mathbf{A}$, respectively. When a single unstable mode dominates, $\lambda_1$ is real and $|\lambda_1|\gg |\lambda_2|$. That is, the leading rank-1 term of $\mathbf{A}$ dominates. Comparing the leading terms of Eq. \eqref{svd_of_e^(At)} and Eq. \eqref{eigen_decomposition_of_e^(At)}, we have
\begin{align}
    \mathbf v_1 \propto \mathbf l_1, \quad \mathbf{C}_{ff} \mathbf u_1 \propto \mathbf r_1.
\end{align}
Thus, the top predictive mode is orthogonal to every eigen-direction except the most unstable direction.

\section{Normalization of output components}\label{app:normalization}
We start from the eigenvalue problems for CCA (Eq. \eqref{eq:cca_eigenproblem} in the main text).
\begin{equation*}
\label{cca}
\begin{aligned}
\mathbf{C}_{pp}^{-1}\mathbf{C}_{pf}\mathbf{C}_{ff}^{-1}\mathbf{C}_{fp}\mathbf{v}_i &= \rho_i^2\,\mathbf{v}_i,\\
\mathbf{C}_{ff}^{-1}\mathbf{C}_{fp}\mathbf{C}_{pp}^{-1}\mathbf{C}_{pf}\mathbf{u}_i &= \rho_i^2\,\mathbf{u}_i.
\end{aligned}
\end{equation*}
It follows that
\begin{equation}
\begin{aligned}
\mathbf G \mathbf Z_p[i]&=\mathbf{P}^\top \mathbf{C}_{pp}^{-1}\mathbf{C}_{pf}\mathbf{C}_{ff}^{-1}\mathbf{F} (\mathbf P^\top \mathbf v_i)\\
&=\mathbf{P}^\top \mathbf{C}_{pp}^{-1}\mathbf{C}_{pf}\mathbf{C}_{ff}^{-1}\mathbf{C}_{fp} \mathbf v_i\\
&=\rho_i^2 \mathbf Z_p[i].
\end{aligned}
\end{equation}
Thus, $\mathbf Z_p[i]$ is a right eigenvector of $\mathbf G$ with eigenvalue $\rho_i^2$. Similarly, $\mathbf Z_f[i]$ is a left eigenvector of $\mathbf G$ with eigenvalue $\rho_i^2$. Notice that
\begin{equation}
   \mathbf Z_p[i]^\top \mathbf Z_f[j]=\mathbb{E}[(\mathbf v_i^\top \mathbf p_t) (\mathbf u_j^\top \mathbf f_t)]=\rho_i\delta_{ij}.
\end{equation}
Therefore, when $k=n$ is the rank of $\mathbf G$, we can eigen-decompose $\mathbf G$ as follows.
\begin{equation}
\begin{aligned}
\mathbf G &=\sum_{i=1}^k \rho_i^2\mathbf Z_p[i]\mathbf Z_f[i]^\top\frac{1}{\rho_i}\\
&=\mathbf{Z}_p^\top \mathbf{R}\mathbf{Z}_f.
\end{aligned}
\end{equation}

\section{Invariance to Affine Changes of Coordinates}\label{app:invariance}
Consider an invertible linear change of coordinates of the state,
$\mathbf{x}'_t = \mathbf{B}\mathbf{x}_t + \mathbf{c}$, with $\mathbf{B}$ an invertible
$n\times n$ matrix and $\mathbf{c}\in\mathbb{R}^n$, and denote by primes all quantities
computed from the transformed trajectory. Temporal differencing annihilates the offset,
\begin{equation}
\delta \mathbf{x}'_t
= \mathbf{x}'_{t+\tau} - \mathbf{x}'_t
= \mathbf{B}(\mathbf{x}_{t+\tau} - \mathbf{x}_t)
= \mathbf{B}\,\delta \mathbf{x}_t ,
\end{equation}
so $\mathbf{P}' = \mathbf{B}\mathbf{P}$, $\mathbf{F}' = \mathbf{B}\mathbf{F}$ and
$\mathbf{C}'_{\alpha\beta} = \mathbf{B}\mathbf{C}_{\alpha\beta}\mathbf{B}^\top$ for
$\alpha,\beta \in \{p,f\}$. Substituting into the predictive similarity matrix of the main
text,
\begin{equation}
\begin{aligned}
\mathbf{G}'
&= \mathbf{P}^\top \mathbf{B}^\top
   \left(\mathbf{B}\mathbf{C}_{pp}\mathbf{B}^\top\right)^{-1}
   \left(\mathbf{B}\mathbf{C}_{pf}\mathbf{B}^\top\right)
   \left(\mathbf{B}\mathbf{C}_{ff}\mathbf{B}^\top\right)^{-1}
   \mathbf{B}\mathbf{F}\\
&= \mathbf{P}^\top
   (\mathbf{B}^\top\mathbf{B}^{-\top})\,\mathbf{C}_{pp}^{-1}\,
   (\mathbf{B}^{-1}\mathbf{B})\,\mathbf{C}_{pf}\,
   (\mathbf{B}^\top\mathbf{B}^{-\top})\,\mathbf{C}_{ff}^{-1}\,
   (\mathbf{B}^{-1}\mathbf{B})\,\mathbf{F}
= \mathbf{G} .
\end{aligned}
\end{equation}
Since the objectives \eqref{eq:unconstrained_objective} and \eqref{eq:non_neg} depend on the
data only through $\mathbf{G}$, the learned outputs $\mathbf{Z}$ are unchanged. The feedforward
weights are not: with $\mathbf{N}=\mathbf{Z}\mathbf{G}$ also unchanged,
$\mathbf{W}'\mathbf{P}' = \mathbf{N}{\mathbf{G}'}^\top = \mathbf{W}\mathbf{P}$, and
$\mathbf{P}' = \mathbf{B}\mathbf{P}$ gives $\mathbf{W}' = \mathbf{W}\mathbf{B}^{-1}$, as also
follows from the update rule for $\mathbf{W}$ in Algorithm~\ref{online_CCA}. Temporal
differencing removes the translation, while whitening by $\mathbf{C}_{pp}$ and
$\mathbf{C}_{ff}$ removes the remaining linear part, leaving the outputs invariant and the
filters equivariant.

Two remarks. The derivation uses only
$\mathbf{C}'_{\alpha\beta}=\mathbf{B}\mathbf{C}_{\alpha\beta}\mathbf{B}^\top$, which holds for
the empirical covariances computed from a single trajectory as well as in the population limit,
so the invariance is exact at finite sample. The numerical implementation replaces the inverses
by Tikhonov-regularized ones, $\mathbf{S}^{+}=(\mathbf{S}+\epsilon\mathbf{I})^{-1}$
(Appendix~\ref{app:numerics}); since $\epsilon\mathbf{I}$ does not transform with the data, the
invariance then holds exactly for orthogonal $\mathbf{B}$ and only approximately for general
invertible $\mathbf{B}$, including pure rescalings $\mathbf{B}=s\mathbf{I}$.

\section{Legendre Transform of Similarity Matching Objective}\label{app:legendre}
We apply the Legendre transform to the similarity matching objective.
\begin{equation}
\label{legendre transform}
\begin{aligned}
&\arg\min_\mathbf{Z} ||\mathbf{G} \mathbf{G}^\top-\mathbf{Z}^\top \mathbf{Z}||^2_F\\
=&\arg\min_\mathbf{Z} [-2\Tr(\mathbf{Z}^\top \mathbf{Z}\mathbf{G}\mathbf{G}^\top)+\Tr(\mathbf{Z}^\top \mathbf{Z} \mathbf{Z}^\top \mathbf{Z})]\\
=& \arg\min_\mathbf{Z} \min_{\mathbf{N}}\max_\mathbf{M} [-4 \Tr(\mathbf{Z}^\top \mathbf{N} \mathbf{G}^\top)+2\Tr(\mathbf{Z}^\top \mathbf{M} \mathbf{Z})\\
&\quad\quad +2\Tr(\mathbf{N}^\top \mathbf{N})-\Tr(\mathbf{M}^\top \mathbf{M})].
\end{aligned}
\end{equation}
To see why the second line and the third line are equal, one can set the derivatives of the last line with respect to $\mathbf N$ and $\mathbf M$ to zero to find the optima
\begin{equation}
    \mathbf{N}^*=\mathbf{Z}\mathbf{G},\quad \mathbf{M}^*=\mathbf{Z}\mathbf{Z}^\top.
\end{equation}
Substituting the optima back into the last line finishes the proof.

\section{Similarity Matching Recovers the Low-Dimensional Principal Subspace}\label{app:subspace}
\begin{figure}
    \centering
\includegraphics[width=1\linewidth]{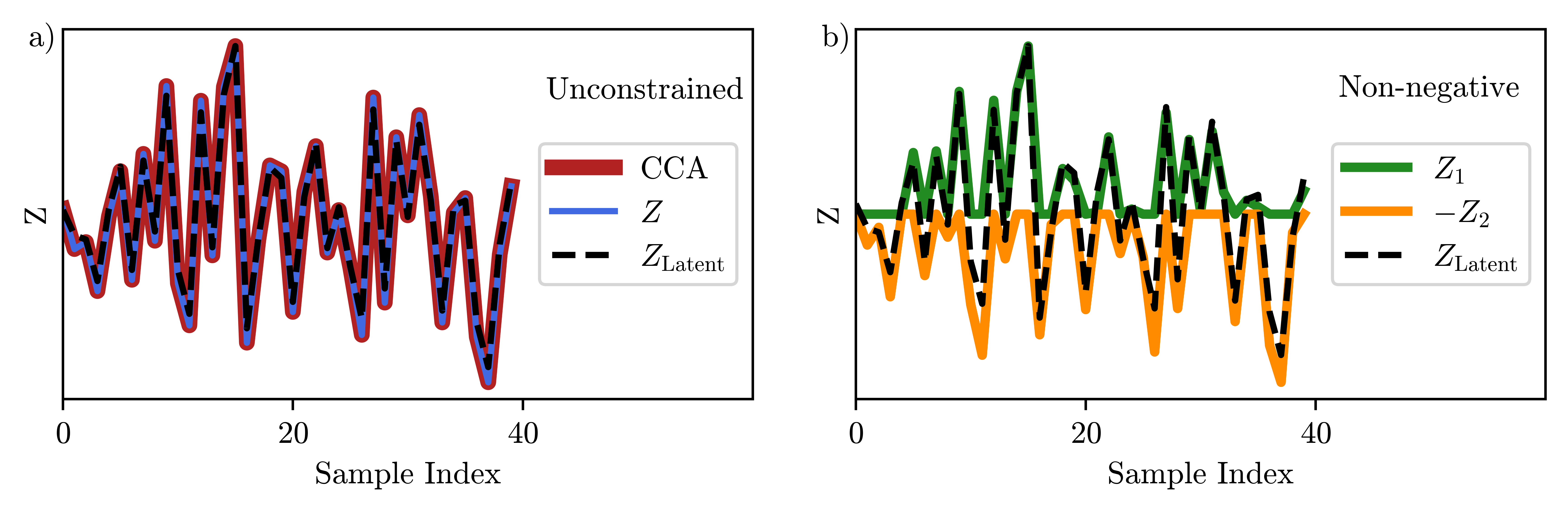}
    \caption{\textbf{Gaussian latent variable.}
    A $1D$ Gaussian latent variable, $\mathbf{Z}_{\text{latent}}$, is projected onto two high-dimensional variables, $\mathbf{X}$ and $\mathbf{Y}$.
    (a) Standard CCA and unconstrained algorithm with $k = 1$ correctly retrieve the latent structure.
    (b) The non-negative algorithm with $k = 2$ learns positive and negative projections of the latent structure.}
    \label{fig:cca_comparison}
\end{figure}

The solution $\mathbf{Z}^*$ obtained from the unconstrained problem
\begin{equation}
\label{eq:similarity_matching_si}
\mathbf{Z}^* = \arg\min_{\mathbf{Z}\in\mathbb{R}^{k\times T}}
\big\|\mathbf{G}\mathbf{G}^\top - \mathbf{Z}^\top\mathbf{Z}\big\|_F^2
\,\text{ is}
\end{equation}
\begin{equation}
\label{eq:solution}
\begin{aligned}
\mathbf{Z}^*
&= \mathbf{M}^{-1}\mathbf{N}\mathbf{G}^\top
= \mathbf{M}^{-1}\mathbf{N}\mathbf{F}^\top \mathbf{C}_{ff}^{-1}\mathbf{C}_{fp}\mathbf{C}_{pp}^{-1}\mathbf{P} \\
&= \mathbf{M}^{-1}\mathbf{W}\mathbf{P}
\simeq \mathbf{V}^\top \mathbf{P},
\end{aligned}
\end{equation}
where $\simeq$ denotes equality up to an orthogonal transformation and component scaling (Section~\ref{sec:lpd}), and recovers the $k$-dimensional subspace spanned by the first $k$
canonical components of the CCA problem between $\mathbf{P}$ and $\mathbf{F}$.
A proof that similarity matching objectives successfully recover the low-dimensional subspace defined by a similarity matrix, such as $\bm{G} \bm{G}^\top$ in Eq.~(\ref{eq:similarity_matching_si}), is provided by \citet{pehlevan2015hebbian}.
Here, we validate this method empirically by applying the unconstrained algorithm defined in Eq.~(\ref{eq:solution}) to a standard benchmark commonly used for evaluating CCA algorithms. We generate synthetic paired datasets $(\mathbf{X}, \mathbf{Y})$ with known latent correlations. Specifically, we consider a one-dimensional latent variable $\mathbf{Z}_{\text{latent}} \in \mathbb{R}^{1 \times d}$ and project it onto high-dimensional spaces via random linear mixing matrices $\mathbf{A}$ and $\mathbf{B}$ to produce
\begin{equation}
\mathbf{X} = \mathbf{Z}_{\text{latent}} \mathbf{A}^\top + \boldsymbol{\epsilon}_X, \quad
\mathbf{Y} = \mathbf{Z}_{\text{latent}} \mathbf{B}^\top + \boldsymbol{\epsilon}_Y,
\end{equation}
where $\boldsymbol{\epsilon}_X$ and $\boldsymbol{\epsilon}_Y$ are Gaussian noise terms.
The algorithm in Eq. \eqref{eq:solution} is applied directly to $(\mathbf{X}, \mathbf{Y})$, without computing finite differences, to demonstrate that the similarity matching solution can successfully recover the underlying low-dimensional subspace, which in this case corresponds to the directions of maximal correlation identified by CCA.
Finally, the canonical variable ($\mathbf{Z}$) computed by the algorithm is compared against the known latent variable $\mathbf{Z}_\text{latent}$. The result is shown in Fig. \ref{fig:cca_comparison}a, also compared with the standard CCA results.
For comparison, Fig.~\ref{fig:cca_comparison}b shows the results of the non-negative algorithm defined in the main text for the constrained problem with $k=2$ output neurons, $\mathbf{Z}_1$ and $\mathbf{Z}_2$.

\section{Whitening and Similarity-Matching Interpretation}\label{app:whitened}
We consider a generic stochastic process with paired variables
\begin{equation*}
\mathbf{p}_t := \mathbf{x}_t,
\qquad
\mathbf{f}_t := \mathbf{x}_{t+1},
\qquad
\mathbf{x}_t \in \mathbb{R}^n.
\end{equation*}
We define the second-order statistics
\begin{equation*}
\mathbf{C}_{pp} = \mathbb{E}[\mathbf{p}_t \mathbf{p}_t^\top],
\qquad
\mathbf{C}_{ff} = \mathbb{E}[\mathbf{f}_t \mathbf{f}_t^\top],
\end{equation*}
and the cross-covariances
\begin{equation*}
\mathbf{C}_{pf} = \mathbb{E}[\mathbf{p}_t \mathbf{f}_t^\top],
\qquad
\mathbf{C}_{fp} = \mathbf{C}_{pf}^\top.
\end{equation*}
All matrices lie in $\mathbb{R}^{n \times n}$.
To isolate interaction structure independently of marginal statistics, we introduce whitened variables
\begin{equation*}
\tilde{\mathbf{p}}_t = \mathbf{C}_{pp}^{-1/2}\mathbf{p}_t,
\qquad
\tilde{\mathbf{f}}_t = \mathbf{C}_{ff}^{-1/2}\mathbf{f}_t,
\end{equation*}
and the corresponding cross-covariance in whitened coordinates
\begin{equation*}
\tilde{\mathbf{C}}_{pf}
=
\mathbf{C}_{pp}^{-1/2}\mathbf{C}_{pf}\mathbf{C}_{ff}^{-1/2},
\qquad
\tilde{\mathbf{C}}_{fp} = \tilde{\mathbf{C}}_{pf}^\top.
\end{equation*}
This transformation removes the influence of marginal covariances, retaining only the interaction structure between past and future variables.
We now consider the matrix
\begin{equation*}
\mathbf{G} =
\mathbf{P}^\top
\mathbf{C}_{pp}^{-1}
\mathbf{C}_{pf}
\mathbf{C}_{ff}^{-1}
\mathbf{F},
\end{equation*}
where $\mathbf{P}$ and $\mathbf{F}$ collect samples of $\mathbf{p}_t$ and $\mathbf{f}_t$ over time.
A direct computation of its Gram matrix yields
\begin{align*}
\mathbf{G}\mathbf{G}^\top
&=
\mathbf{P}^\top
\mathbf{C}_{pp}^{-1}
\mathbf{C}_{pf}
\mathbf{C}_{ff}^{-1}
\mathbf{F}\mathbf{F}^\top
\mathbf{C}_{ff}^{-1}
\mathbf{C}_{fp}
\mathbf{C}_{pp}^{-1}
\mathbf{P}.
\end{align*}
For long trajectories, we use the empirical approximation
\begin{equation*}
\mathbf{F}\mathbf{F}^\top \approx T\,\mathbf{C}_{ff},
\end{equation*}
which gives
\begin{align*}
\mathbf{G}\mathbf{G}^\top
&=
T\,\mathbf{P}^\top
\mathbf{C}_{pp}^{-1}
\mathbf{C}_{pf}
\mathbf{C}_{ff}^{-1}
\mathbf{C}_{fp}
\mathbf{C}_{pp}^{-1}
\mathbf{P}.
\end{align*}
Rewriting in whitened coordinates yields
\begin{equation*}
\mathbf{G}\mathbf{G}^\top
=
T\,\tilde{\mathbf{P}}^\top
\tilde{\mathbf{C}}_{pf}
\tilde{\mathbf{C}}_{pf}^\top
\tilde{\mathbf{P}}.
\end{equation*}
This shows that the relevant structure is fully captured by the whitened cross-covariance acting on whitened inputs.
Defining
\begin{equation*}
\bar{\mathbf{X}} := \tilde{\mathbf{C}}_{fp}\tilde{\mathbf{P}},
\end{equation*}
we obtain the compact similarity form
\begin{equation*}
\mathbf{G}\mathbf{G}^\top = \bar{\mathbf{X}}^\top \bar{\mathbf{X}},
\end{equation*}
which connects this construction to standard similarity-matching objectives.
\textbf{Special case: linear dynamics.}
If the dynamics are linear,
\begin{equation*}
\mathbf{f}_t = \mathbf{A}\mathbf{p}_t + \mathbf{w}_t,
\end{equation*}
then the cross-covariance satisfies
\begin{equation*}
\mathbf{C}_{fp} = \mathbf{A}\mathbf{C}_{pp}.
\end{equation*}
Substituting into the whitened expression gives
\begin{align*}
\tilde{\mathbf{C}}_{fp}
&=
\mathbf{C}_{ff}^{-1/2}
\mathbf{C}_{fp}
\mathbf{C}_{pp}^{-1/2} \\
&=
\mathbf{C}_{ff}^{-1/2}
\mathbf{A}
\mathbf{C}_{pp}^{1/2}
=: \tilde{\mathbf{A}}.
\end{align*}
Note that in whitened coordinates, the dynamics reduce to
\begin{equation*}
\tilde{\mathbf{f}}_t = \tilde{\mathbf{A}}\tilde{\mathbf{p}}_t.
\end{equation*}

\section{Details on Numerical Implementation of LPD}\label{app:numerics}
The offline implementation of LPD solves the min--max problem of Eq.~\eqref{eq:min-max legend} by alternating gradient updates: an inference step that iterates a rectified update of $\mathbf{Z}$ at fixed $\mathbf{W},\mathbf{M}$, followed by a learning step that updates $\mathbf{W}$ and $\mathbf{M}$ from empirical covariance estimates (Algorithm~\ref{Aoffline}), where $[\cdot]_+$ denotes elementwise rectification. The covariance matrices are defined as
\begin{equation}
    \mathbf{C}_{zp}:=\mathbf{Z}\mathbf{P}^\top,
    \qquad
    \mathbf{C}_{zz}:=\mathbf{Z}\mathbf{Z}^\top,
\end{equation}
and similarly for the remaining terms. Regularization is applied through Tikhonov-regularized inverses, $\mathbf{S}^{+}:=(\mathbf{S}+\epsilon \mathbf{I})^{-1}$, to ensure numerical stability.
\begin{algorithm2e}[h]
\caption{Learning Local Predictive Directions Offline}
\label{Aoffline}
\KwIn{past and future increment matrices $\mathbf{P},\mathbf{F}$; number of output units $k$}
Initialize $\mathbf{W} \in \mathbb{R}^{k \times n}$ and $\mathbf{M} \in \mathbb{R}^{k \times k}$ with $\mathbf{M} \succ 0$\;
\Repeat{relative changes in $\mathbf{W}$ and $\mathbf{M}$ fall below threshold}{
  \textbf{Inference:} update $\mathbf{Z}$ for fixed $\mathbf{W},\mathbf{M}$\;
  \Repeat{inference convergence}{
    $\mathbf{Z} \leftarrow [\mathbf{Z} + \eta_Z(\mathbf{W}\mathbf{P} - \mathbf{M}\mathbf{Z})]_+$\;
  }
  \textbf{Learning:} update empirical covariances
  $\mathbf{C}_{pp},\mathbf{C}_{ff},\mathbf{C}_{pf},\mathbf{C}_{fp},\mathbf{C}_{zp},\mathbf{C}_{zz}$\;
  $\mathbf{W} \leftarrow \mathbf{W} + \eta_W
   \left(
     \mathbf{C}_{zp} \mathbf{C}_{pp}^{+} \mathbf{C}_{pf} \mathbf{C}_{ff}^{+}
     \mathbf{C}_{fp} \mathbf{C}_{pp}^{+} - \mathbf{W}
   \right)$\;
  $\mathbf{M} \leftarrow \mathbf{M} + \eta_M(\mathbf{C}_{zz}-\mathbf{M})$\;
}
\end{algorithm2e}
Convergence of the offline algorithm is assessed by monitoring the relative parameter changes
\begin{equation}
    \frac{\|\mathbf{W}^{(r+1)}-\mathbf{W}^{(r)}\|_F}{\|\mathbf{W}^{(r)}\|_F},
    \qquad
    \frac{\|\mathbf{M}^{(r+1)}-\mathbf{M}^{(r)}\|_F}{\|\mathbf{M}^{(r)}\|_F},
\end{equation}
and terminating the iterations once both quantities fall below a predefined threshold.
Both the online and offline implementations of the algorithm can be run on a single standard GPU for all experiments reported in the paper.
Moreover, we note that applying the offline algorithm to the same data reproduces the results obtained with the online algorithm, with no significant differences. More generally, when the dataset is sufficiently large and the underlying dynamical process is stationary, the online and offline implementations yield equivalent results.

\section{Details on stochastic dynamical systems}\label{app:dynamical_systems}
\subsection{Affine Saddle-Point Dynamics --- Fig.~\ref{fig:dynamical_systems}a of the Main Text}\label{app:affine}

Let $\mathbf{x}_t \in \mathbb{R}^2$ denote the system state at time $t$, and let $\mathbf{x}^\ast \in \mathbb{R}^2$ represent the saddle point. The dynamics are given by
\begin{equation}
\mathbf{x}_{t+1} = \mathbf{x}^\ast + \mathbf{A} \, (\mathbf{x}_t - \mathbf{x}^\ast) + \sigma \, \boldsymbol{\xi}_t,
\qquad \boldsymbol{\xi}_t \sim \mathcal{N}(\mathbf{0}, \mathbf{I}),
\end{equation}
where $\mathbf{A} = \mathrm{diag}(\lambda_{\mathrm{u}}, \lambda_{\mathrm{s}})$ is a diagonal matrix with one unstable eigenvalue $\lambda_{\mathrm{u}} > 1$ and one stable eigenvalue $0 < \lambda_{\mathrm{s}} < 1$, and $\sigma$ controls the level of stochasticity. In this case the unstable direction is along the $x_1$ axis, $\mathbf{d}_{\mathrm{unstable}} = (1, 0)$.

\subsection{Generalized Lotka--Volterra Systems --- Fig.~\ref{fig:dynamical_systems}b of the Main Text}

To validate our algorithm in the presence of multiple saddle points, we apply it to trajectories generated by a stochastic population dynamics model known as the generalized Lotka--Volterra (gLV) system, commonly used to illustrate stable heteroclinic channels and winnerless competition:
\begin{equation}
\label{eq:glv}
\dot{x}_i = x_i \left( 1 - \sum_j A_{ji} \, x_j \right) + \sigma \, \xi_i(t),
\qquad \xi_i(t) \sim \mathcal{N}(0, 1),
\end{equation}
where $x_i$ denotes the abundance of species $i$, $A_{ij}$ encodes the effect of species $i$ on species $j$, and $\sigma$ sets the noise amplitude. A positive $A_{ij}$ indicates that species $i$ has a detrimental effect on species $j$; if $A_{ji}$ is also positive, the two species interact competitively. By tuning these interactions, the system can enter regimes exhibiting stable heteroclinic channels~\citep{bick2010occurrence, afraimovich2008winnerless}.

A convenient set of saddle points can be constructed by considering fixed points in which a single species is active ($x_i > 0$) while all others are extinct ($x_j = 0$ for $j \neq i$). Setting the self-interactions to $A_{ii} = 1$ for all species $i$ places these fixed points at $x_i = 1$, so that linearizing Eq.~\eqref{eq:glv} about the saddle dominated by species $i$ gives $\dot{x}_j \approx (1 - A_{ij}) \, x_j$ for every $j \neq i$. Each row of the interaction matrix therefore collects the transverse eigenvalues around the corresponding equilibrium: entries with $A_{ij} < 1$ generate unstable directions, guiding the system from the saddle dominated by species $i$ toward the saddle where species $j$ becomes active.

We study a system with $N = 3$ species arranged to form a stable heteroclinic channel (SHC), producing sequential transitions between saddle points. Transitions from the saddle where species $i$ is active to the saddle where species $j$ becomes active are induced by setting $A_{ij} = 0.7$, while all other off-diagonal entries of the interaction matrix are set to $1.5$.

\subsection{Scalability in High-Dimensional Generalized Lotka--Volterra Systems} \label{app:high_dimensional_glv}

We evaluate the scalability of LPD with respect to system dimensionality using generalized Lotka--Volterra (gLV) dynamics. We consider varying system sizes $N \in \{3, 5, 10, 25, 50, 100\}$, across noise levels $\sigma \in \{0.01, 0.05, 0.1\}$.

For each configuration, we simulate trajectories and apply LPD to learn predictive representations. The generalized Lotka--Volterra dynamics are given by
\begin{equation}
\dot{\bm{x}} = \bm{x} \odot \left( \bm{1} - \bm{A}^\top\bm{x} \right) + \sigma \,\bm{\xi}(t),
\end{equation}
where $\bm{A} \in \mathbb{R}^{N \times N}$ induces sequential switching dynamics (same convention as Eq.~\eqref{eq:glv}) and $\bm{\xi}(t)$ is Gaussian white noise. Simulations are performed using Euler integration with $\mathrm{d}t = 0.1$ over $T=5000$, starting near single-species dominance with small perturbations. After removing an initial transient of $10^3$ steps, we construct $N_{\mathrm{data}} = 19{,}000$ training samples.

Performance is evaluated by comparing the inferred transition structure to the ground-truth transition graph induced by $\bm{A}$. Since components are unordered, we align learned and ground-truth transitions using the Hungarian algorithm applied to a similarity matrix between normalized rows of $\bm{M}^{-1}\bm{W}$ and the theoretical transition matrix. The resulting agreement is quantified using NMI.

We observe that LPD remains robust as system dimensionality increases, with consistently high NMI across all tested values of $N$, without requiring any re-tuning of optimization hyperparameters, and across all noise levels.

\begin{figure}[h]
    \centering
    \includegraphics[width=0.7\linewidth]{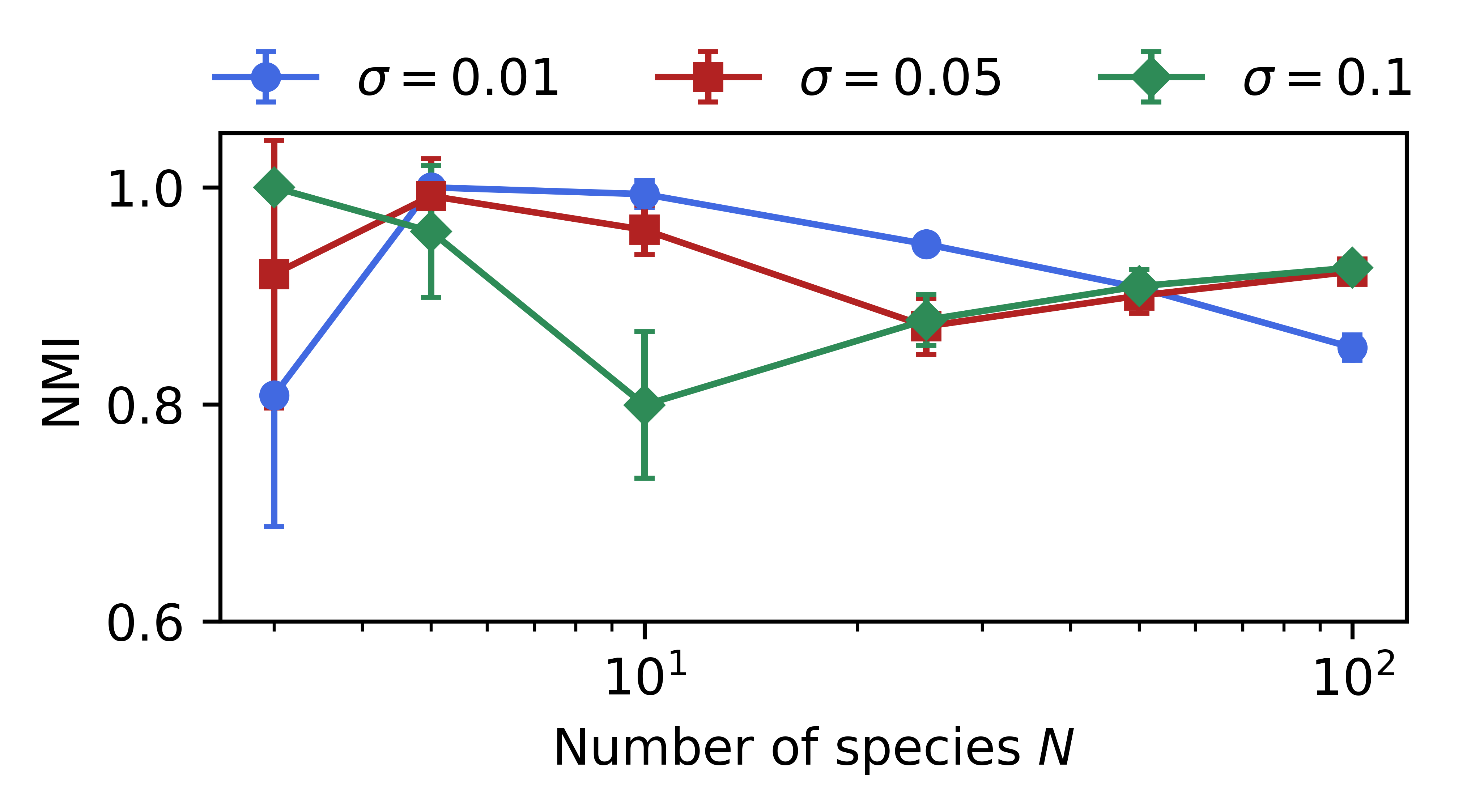}
    \caption{\textbf{Scalability in gLV systems.} Normalized Mutual Information (NMI) as a function of system size $N$ under different noise levels, reported as mean $\pm$ standard deviation over $10$ realizations. Results are for the offline implementation Alg. \ref{Aoffline}.}
    \label{fig:robustness_glv}
\end{figure}

\subsection{A Sphere with Two Saddle Points --- Fig.~\ref{fig:dynamical_systems}c of the Main Text}
To further test the algorithm on a nonlinear, curved manifold, we consider a two-dimensional flow on the surface of a sphere. Flows on the sphere defined by stream functions in spherical coordinates are a standard tool in dynamical systems and fluid mechanics and naturally exhibit heteroclinic dynamics~\citep{peixoto1962structural, batchelor2000introduction}.
We consider stochastic dynamics constrained to the surface of the unit sphere, where the state $\mathbf{x}(t) \in \mathbb{R}^3$ evolves under a combination of deterministic flow and isotropic noise tangent to the sphere. The deterministic component is derived from a stream function
\begin{equation}
\psi(\theta, \phi) = \sin^2 \theta \, \cos(2\phi),
\end{equation}
defined in spherical coordinates $(\theta,\phi)$, with $\theta$ the polar angle from the positive $z$-axis and $\phi$ the azimuthal angle in the $xy$-plane. The gradient of the stream function defines a divergence-free flow on the sphere, with velocity components
\[
v_\theta = -\frac{1}{\sin\theta}\frac{\partial \psi}{\partial \phi}, \quad
v_\phi = \frac{\partial \psi}{\partial \theta}
\]
along the local tangent vectors $\mathbf{e}_\theta$ and $\mathbf{e}_\phi$. Stochastic perturbations are added as Gaussian noise projected onto the tangent plane via the projection matrix $\mathbf    \Pi = \mathbf I - \mathbf{x}\mathbf{x}^\top$, ensuring that the dynamics remain on the sphere. The resulting stochastic differential equation is
\[
d\mathbf{x} = \mathbf{v}(\mathbf{x})\,dt + \sigma \mathbf \Pi (\mathbf{x})\,d\mathbf{w}_t,
\]
where $\mathbf{v}(\mathbf{x}) = v_\theta \mathbf{e}_\theta + v_\phi \mathbf{e}_\phi$ and $\sigma$ controls the noise amplitude.
Trajectories are integrated numerically with an Euler--Maruyama scheme, followed by renormalization to maintain the unit-norm constraint. This formulation captures both structured, divergence-free flows on curved manifolds and stochastic variability confined to the manifold.
This system contains two saddle points located at the north ($\theta = 0$) and south ($\theta = \pi$) poles of the sphere. The unstable manifolds correspond to directions in the tangent plane along which trajectories locally depart from these poles, as determined by linearizing the velocity field around the fixed points:
\begin{itemize}
    \item \textbf{North pole ($\theta = 0$):} unstable directions along $(1, 1)$ and $(-1, -1)$ in the tangent plane spanned by $x_1$ and $x_2$ at $x_3 = 1$.
    \item \textbf{South pole ($\theta = \pi$):} unstable directions along $(1, -1)$ and $(-1, 1)$ in the tangent plane spanned by $x_1$ and $x_2$ at $x_3 = -1$.
\end{itemize}

\section{Model Selection and Validation}\label{app:model_selection}
\begin{figure}[h]
    \centering
    \includegraphics[width=1\linewidth]{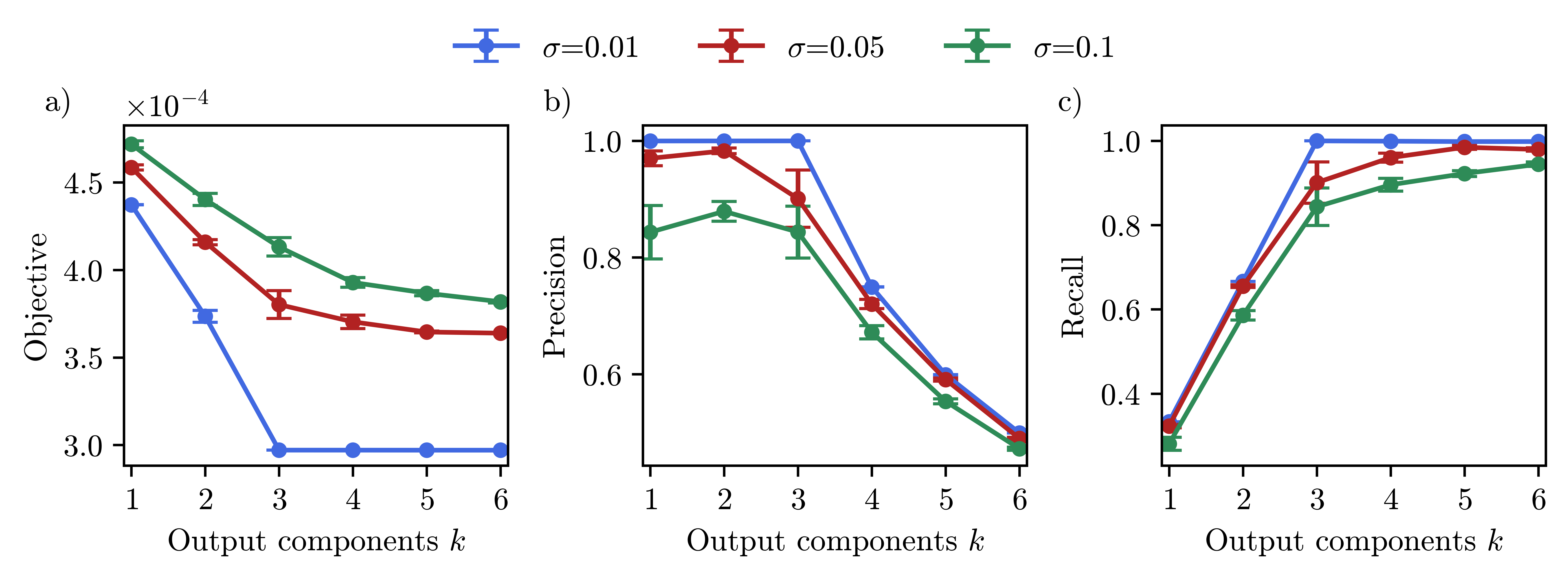}
    \caption{\textbf{Model selection and validation for Section~\ref{sec:dynamical_systems_exp} (SHC in gLV system).} Results are shown as mean $\pm$ standard error over $10$ trials. (a) Objective function as a function of the number of components $k$ across noise levels. (b--c) Precision and recall (Eq.~\eqref{eq:precision_recall}) across $k$ and noise levels.}
    \label{fig:model_selection_validation_SI}
\end{figure}

The proposed normative framework provides an internal, task-aligned criterion for selecting the number of output components $k$, evaluated using the same objective optimized during learning rather than an external criterion such as $k$-means clustering or hidden Markov models. Because the objective is non-increasing in $k$, the plateau is a heuristic rather than an explicit complexity penalty; we therefore validate it below against ground-truth directions. In practice, the algorithm is run for a range of candidate values of $k$, and the corresponding objective is evaluated. Once $k$ reaches the intrinsic dimensionality of the unstable structure, the objective exhibits a clear plateau; the resulting elbow provides a natural estimate of the number of dynamical modes (Fig.~\ref{fig:model_selection_validation_SI}a).
We further validate this criterion by computing precision and recall of the alignment between learned and ground-truth directions. These results confirm the expected behavior: for $k < k_{\mathrm{true}}$, the method recovers correct directions but misses components, whereas for $k > k_{\mathrm{true}}$, all true directions are recovered but redundant or spurious components are introduced.
To quantify model selection more precisely, we evaluate the objective
\begin{equation}
   \frac{1}{T^2} \big\|\mathbf{G}\mathbf{G}^\top - \mathbf{Z}^\top \mathbf{Z}\big\|_F^2,
   \label{eq:objective}
\end{equation}
where $T$ is the number of samples used to compute $\mathbf{G}$ (matching the sample size of $\mathbf{Z}$). Since the derivation of $\mathbf{Z}$ neglects constant terms (see Appendix \ref{app:normalization}), these must be reinstated when evaluating the objective. We do so by rescaling each row of $\mathbf{Z}$ with a constant factor, introducing $k$ free parameters, which are optimized to obtain a corrected estimate of Eq.~(\ref{eq:objective}).
To further validate the selection of $k$, we compute matching-based precision and recall between learned directions and ground-truth unstable directions. Let
\begin{equation}
    \{\mathbf{v}_i\}_{i=1}^{k} \quad \text{and} \quad \{\mathbf{d}_j\}_{j=1}^{k_{\mathrm{true}}}
\end{equation}
denote the learned and ground-truth directions, respectively, each normalized to unit norm, $\lVert \mathbf{v}_i \rVert = \lVert \mathbf{d}_j \rVert = 1$.
We define the similarity matrix
\begin{align}
C_{ij} = \max\!\left( \langle \mathbf{v}_i, \mathbf{d}_j \rangle,\, 0 \right),
\end{align}
and compute the optimal one-to-one matching
\begin{align}
\mathcal{M} = \arg\max_{\text{1--1 matchings}} \sum_{(i,j)\in\mathcal{M}} C_{ij}.
\end{align}
Precision and recall are then defined as
\begin{align}
\label{eq:precision_recall}
\mathrm{Precision} &= \frac{1}{k} \sum_{(i,j)\in\mathcal{M}} C_{ij}, &
\mathrm{Recall} &= \frac{1}{k_{\text{true}}} \sum_{(i,j)\in\mathcal{M}} C_{ij}.
\end{align}
Precision quantifies the quality of learned directions while penalizing spurious components, whereas recall measures how many ground-truth directions are recovered. Together, these metrics characterize the effect of under- and over-estimating $k$.

\subsection{Model Selection for Naturalistic Stimuli}
\label{app:modelsel_stimuli}
\begin{figure}[!h]
  \begin{center}
\centerline{\includegraphics[width=0.5\columnwidth]{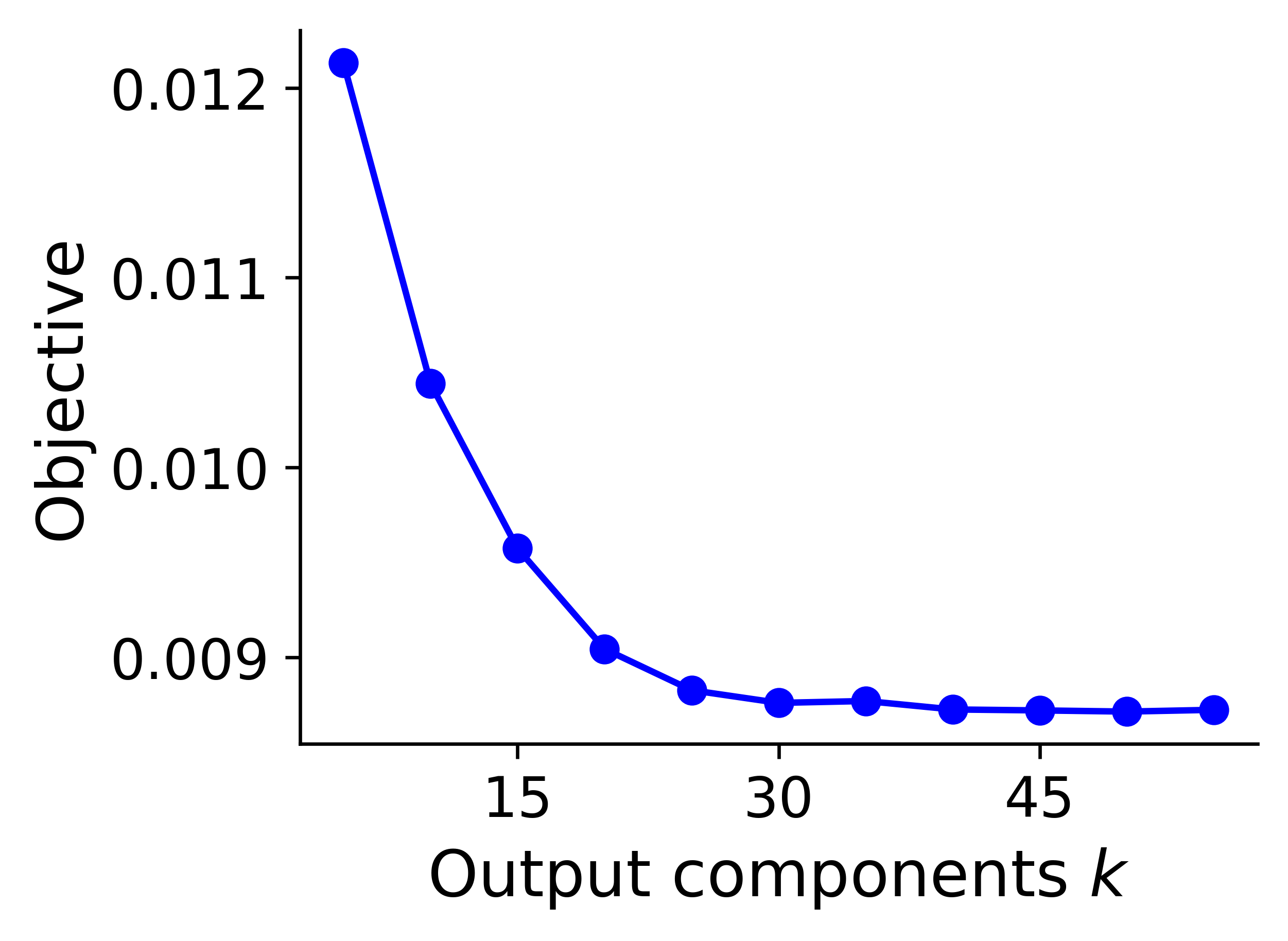}}
    \caption{\textbf{Model selection for the dead-leaves stimulus ensemble.}
    Objective \eqref{eq:objective} as a function of the number of output components $k$,
    for LPD applied to past--future temporal differences of the one-dimensional dead-leaves
    model of Section~\ref{sec:natural_stimuli}. The elbow provides the estimate of the number of
    predictive components used to learn the filters in Fig.~\ref{fig:receptive_fields}.}
    \label{fig:modelsel_stimuli}
  \end{center}
\end{figure}

The same criterion applies to the naturalistic stimulus ensemble of
Section~\ref{sec:natural_stimuli}. We run LPD on past--future temporal-difference pairs obtained
from the one-dimensional dead-leaves model and evaluate the objective \eqref{eq:objective} over a
range of candidate $k$ (Fig.~\ref{fig:modelsel_stimuli}). As for the dynamical systems of
Section~\ref{sec:dynamical_systems_exp}, the objective decreases steeply while additional
components still capture distinct predictive structure, and exhibits a plateau once $k$ exceeds
the intrinsic dimensionality of the ensemble; the elbow again provides a natural estimate of the
number of predictive components, which is the criterion used to select the number of output units
for the filters reported in Fig.~\ref{fig:receptive_fields}. Unlike the generalized
Lotka--Volterra case, this stimulus ensemble has no ground-truth set of predictive directions, so
the precision and recall of Eq.~\eqref{eq:precision_recall} cannot be evaluated and the objective
curve is the only available criterion.

\section{Robustness to Finite-Difference Lag $\tau$}
We assess the sensitivity of LPD to the finite-difference lag $\tau$, which controls the temporal resolution of the tangent-space approximation. For a fixed system size ($N=10$), we construct finite differences as
\begin{equation}
\delta \bm{x}_t = \bm{x}_{t+\tau} - \bm{x}_t,
\end{equation}
with $\tau \in \{1, 5, 10, 15, 20, 25\}$, and repeat the full learning and evaluation pipeline across different noise levels.
Performance is measured using normalized mutual information (NMI) between the inferred and ground-truth transition structures. We find that performance is only weakly affected by the choice of $\tau$, as long as it remains below the characteristic timescales of the system dynamics. This indicates that LPD is robust to moderate changes in temporal discretization.
\begin{figure}[h]
    \centering
    \includegraphics[width=0.5\linewidth]{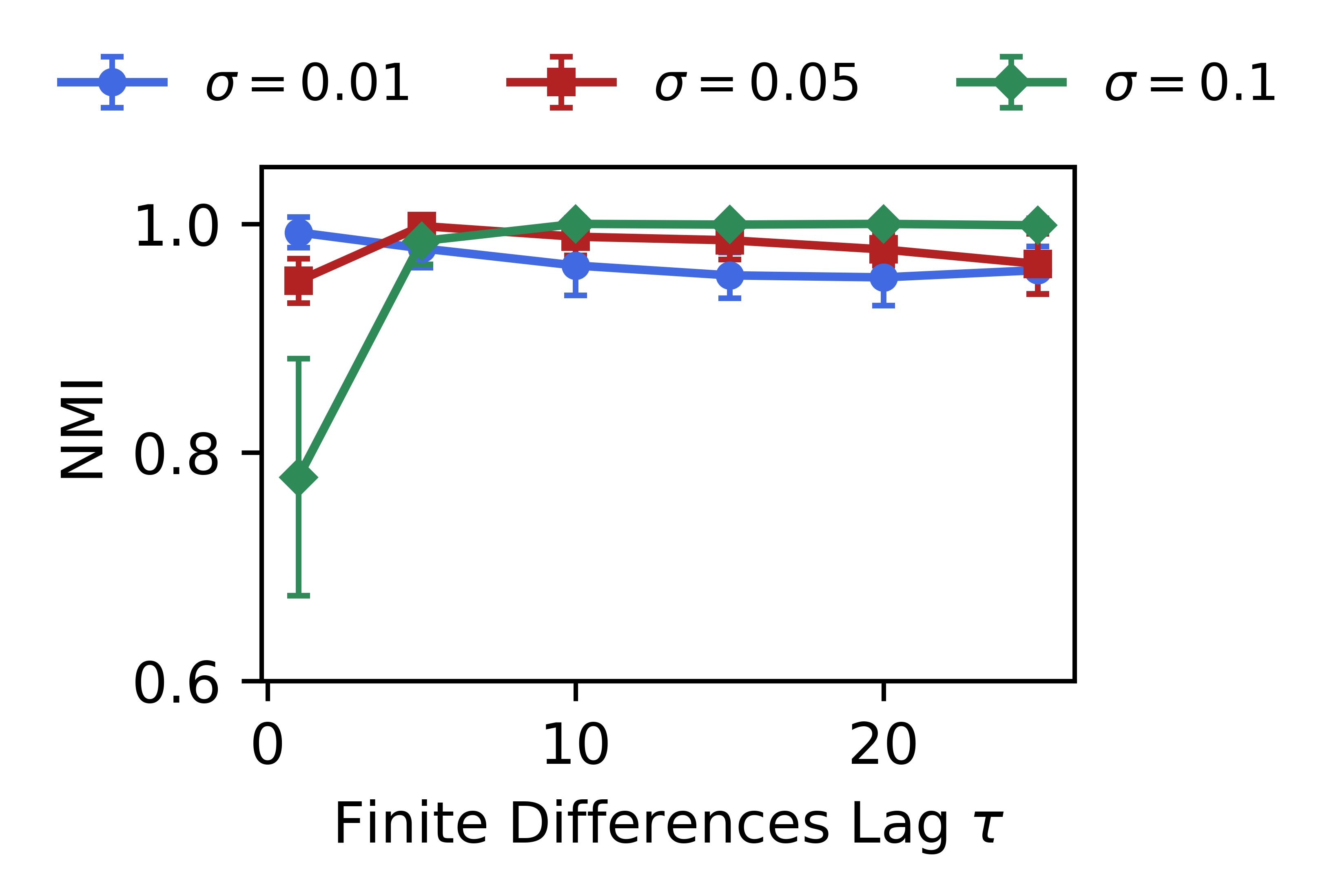}
    \caption{\textbf{Robustness to finite-difference lag.} NMI as a function of lag $\tau$ for $N=10$, across different noise levels. Results are reported as mean $\pm$ standard deviation over $10$ independent trials.}
    \label{fig:robustness_tau}
\end{figure}

\section{Ablation Studies: Finite Differences and Non-Negativity}\label{app:ablation}
\begin{figure}[h]
    \centering
    \includegraphics[width=1\linewidth]{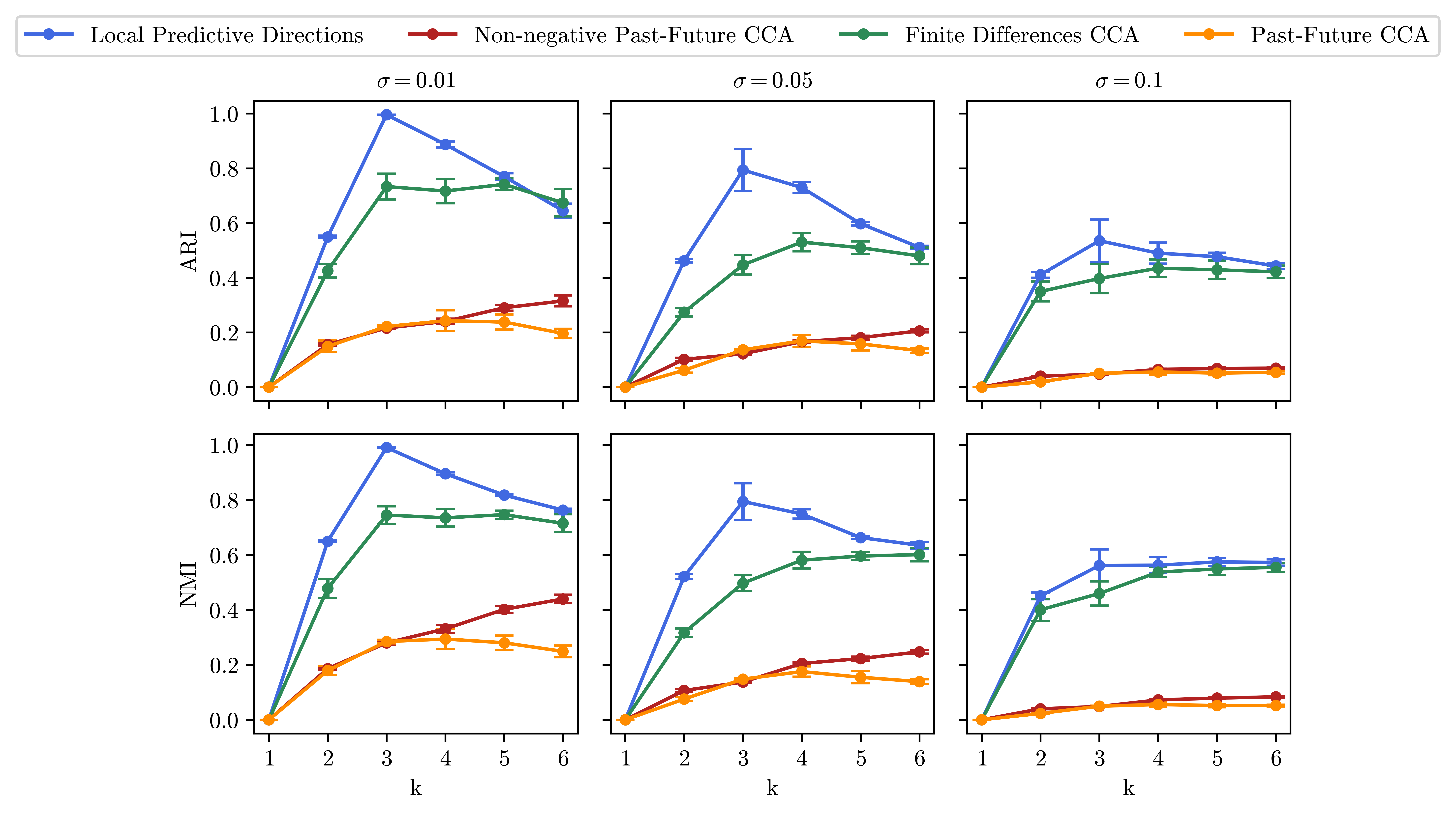}
    \caption{\textbf{Clustering performance across methods.} Adjusted Rand Index (ARI, top) and Normalized Mutual Information (NMI, bottom) as a function of $k$ under different noise levels. Each curve corresponds to a specific combination of non-negativity (ReLU) and finite-difference preprocessing, i.e., PF, NN-PF, FD, or LPD. Results are reported as mean $\pm$ standard deviation over $10$ independent realizations.}
    \label{fig:clustering_performance}
\end{figure}

We study the contribution of finite differences and the non-negativity constraint in LPD by comparing against three baselines: standard past--future CCA (PF), non-negative PF (NN-PF), and unconstrained finite-differences CCA (FD). All models are trained on gLV trajectories (Fig. \ref{fig:dynamical_systems}b) and evaluated on independent test trajectories with matching noise. Ground-truth labels correspond to the next visited saddle, while predictions are obtained from the maximal activation of each model's output. Performance is measured using normalized mutual information (NMI). As shown in Fig.~\ref{fig:clustering_performance}, LPD achieves higher NMI at low and intermediate noise, converging with FD only at high noise, indicating that both finite differences and non-negativity are essential for optimal performance. We also evaluate clustering performance using Adjusted Rand Index (ARI) and NMI across noise levels and number of components $k$.
LPD combines two key ingredients: finite differences for translational invariance and a non-negativity constraint. We compare against PF, NN-PF, and FD baselines under varying $k$ and noise levels.
For evaluation, we segment trajectories into ground-truth labels corresponding to the next saddle visited in the SHC dynamics (gLV system). Each time point is assigned the label of the next saddle based on alignment of the finite difference
\[
\delta \mathbf{x}_t = \mathbf{x}_{t+\tau} - \mathbf{x}_{t},\quad \tau=1,
\]
with known unstable directions. Predicted labels are obtained by selecting the maximal component of each output vector $\mathbf{z}_t$. Performance is quantified using ARI and NMI, measuring clustering agreement with ground truth.

\section{Comparison with Soft Clustering in Tangent Space}\label{app:clustering}
\begin{figure} [h]
    \centering
    \includegraphics[width=0.8\linewidth]{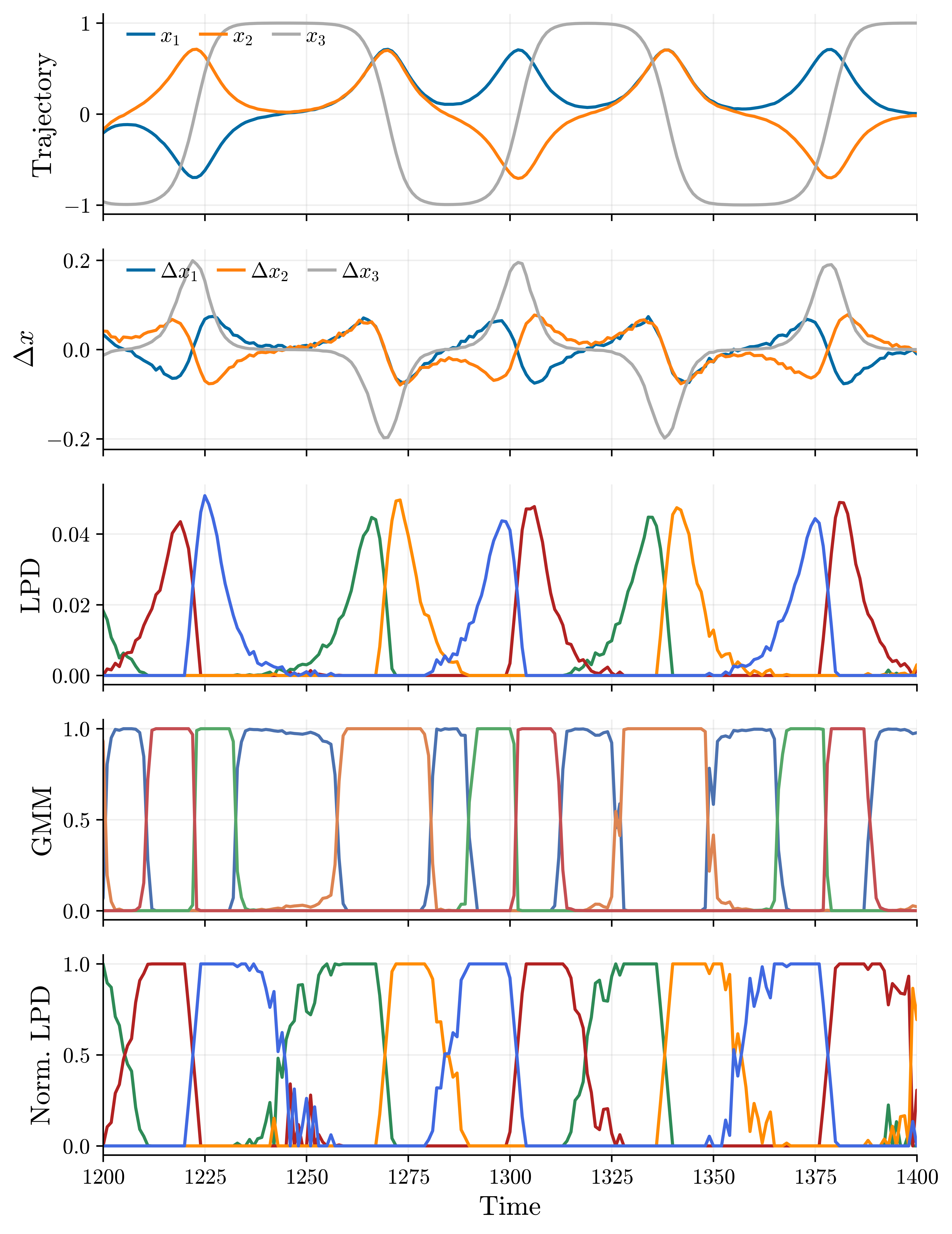}
    \caption{Example trajectories of the dynamical system, corresponding finite-difference features, and learned representations. We show embeddings learned by LPD, soft assignments from a Gaussian Mixture Model (GMM) in tangent space, and normalized LPD obtained by enforcing simplex normalization at each time step. This normalization makes the LPD representation directly comparable to probabilistic soft assignments.
}
    \label{fig:clustering_trajectories}
\end{figure}

\begin{figure}[h]
    \centering
    \includegraphics[width=0.8\linewidth]{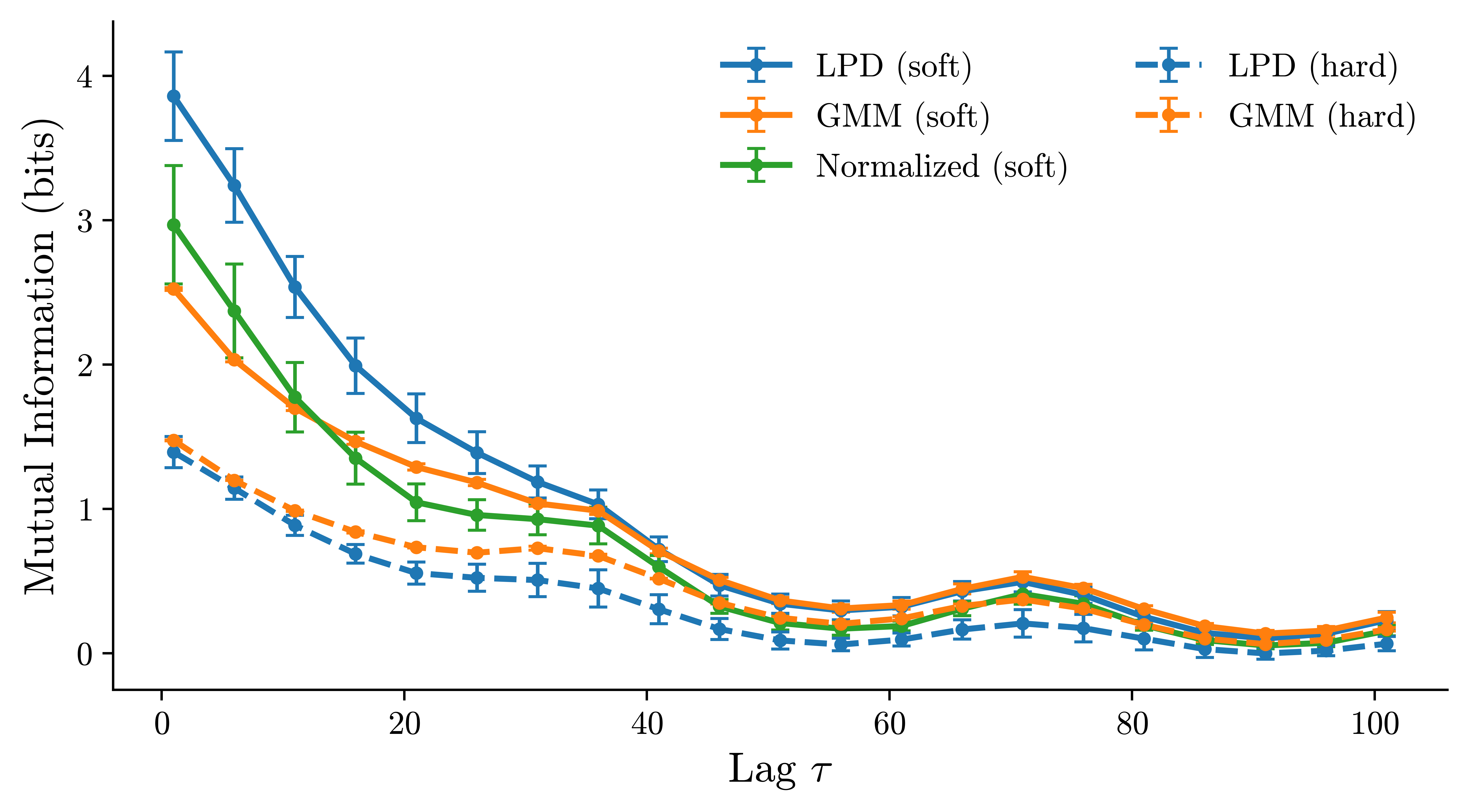}
   \caption{ Time-lagged mutual information between learned representations and future finite-difference dynamics as a function of the prediction horizon $h$.
We compare LPD, GMM, normalized LPD, and their corresponding hard assignments obtained via argmax discretization.
Error bars indicate standard deviation across $10$ independent runs.
}
    \label{fig:MI_comparison}
\end{figure}

We compare LPD with Gaussian Mixture Models (GMMs) in tangent space to assess whether LPD captures predictive structure beyond standard unsupervised clustering. Both methods operate on identical finite-difference features. In particular, for each method we learn a representation $\mathbf{Z}(t)$. We run the experiment on the sphere with two-saddle dynamical system (Fig.~\ref{fig:dynamical_systems}c) with noise level $\sigma = 0.01$. LPD produces embeddings $\mathbf{z}_{\mathrm{LPD}} \in \mathbb{R}^k$, while GMM yields soft assignments $\mathbf{z}_{\mathrm{GMM}}$ with $k=4$ components.
In Fig.~\ref{fig:clustering_trajectories}, we show an example of trajectories of the dynamical system, the corresponding finite differences, the representations learned by LPD, and the soft assignments learned by the Gaussian Mixture Model in tangent space. Finally, we also show the normalized LPD, where at each time step we enforce the components of $\mathbf{Z}$ to sum to $1$, matching the normalization constraint of GMM.
We evaluate the predictive information captured by the learned representations by computing the time-lagged mutual information
\[
I(\mathbf{z}_t ; \delta\mathbf{x}_{t+h}),
\]
as a function of the prediction horizon $h$, where $\delta\mathbf{x}_{t} = \mathbf{x}_{t+\tau}-\mathbf{x}_{t}$ with $\tau=1$.
The results are shown in Fig.~\ref{fig:MI_comparison}, comparing LPD, GMM, normalized LPD, and their corresponding hard assignments obtained by taking the argmax component at each time step.
The results show that LPD captures more predictive information at small lags, while normalized LPD exhibits performance comparable to the Gaussian Mixture Model, as well as to the hard clustering obtained from both LPD and GMM.
In particular, we emphasize that LPD is able to capture additional information because it also learns the magnitude of the finite differences, whereas GMM and normalized LPD are constrained by their normalization conditions. At the north and south poles, for example, all output components are close to zero; nevertheless, the neighborhood retains information about which of the two poles the trajectory is near, due to distinct activation patterns that either drive the dynamics toward or away from each saddle.
GMM, instead, classifies the north and south poles in the same way, since it only depends on instantaneous values of the finite differences and does not account for the temporal correlations that connect these states to their past and future evolution.
\end{document}